\documentclass[prb, superscriptaddress, showkeys, twocolumn]{revtex4-1}
\usepackage{natbib}
\usepackage{graphicx}
\usepackage{dcolumn}
\usepackage{amsmath}
\usepackage{amssymb}
\usepackage{subfigure}
\usepackage[dvipsnames]{xcolor}
\usepackage{bm}
\usepackage{siunitx}

\newcommand{\bp}{{\bf p}}
\newcommand{\br}{{\bf r}}
\newcommand{\bn}{{\bf n}}
\newcommand{\bnz}{{\bf n}_{\zeta}}
\newcommand{\ba}{{\bf a}}
\newcommand{\bs}{{\bf s}}
\newcommand{\bj}{{\bf j}}
\newcommand{\bz}{\hat{\bf z}}
\newcommand{\bsigma}{\boldsymbol{\sigma}}
\newcommand{\bcalA}{\boldsymbol{\mathcal{A}}}
\newcommand{\bcalF}{\boldsymbol{\mathcal{F}}}
\newcommand{\betasf}{\beta_s}
\newcommand{\betaDP}{\beta_\mathrm{DP}}
\newcommand{\betatau}{\beta_\tau}
\newcommand{\tausf}{\tau_s}
\newcommand{\tauDP}{\tau_\mathrm{DP}}
\newcommand{\Exc}{\Delta_\mathrm{xc}}
\newcommand{\lsf}{\ell_s}
\def\be{\begin{equation}}
\def\ee{\end{equation}}
\def\ber{\begin{eqnarray}}
\def\eer{\end{eqnarray}}

\begin{document}
\title{Spin-charge coupled dynamics driven by a time-dependent magnetization}
\author{Sebastian T\"{o}lle}
\affiliation{Universit\"at Augsburg, Institut f\"ur Physik, 86135 Augsburg, Germany}
\author{Ulrich Eckern}
\affiliation{Universit\"at Augsburg, Institut f\"ur Physik, 86135 Augsburg, Germany}
\author{Cosimo Gorini}
\affiliation{Universit\"at Regensburg, Institut f\"ur Theoretische Physik, 93040 Regensburg, Germany}

%\date{\today}
%\pacs{72.25.-b}

\begin{abstract}
The spin-charge coupled dynamics in a thin, magnetized metallic system are investigated.
The effective driving force acting on the charge carriers is generated by a dynamical magnetic
texture, which can be induced, e.g., by a magnetic material in contact with a normal-metal system.  
We consider a general inversion-asymmetric substrate/normal-metal/magnet structure,
which, by specifying the precise nature of each layer, can mimick various experimentally employed setups.
Inversion symmetry breaking gives rise to an effective Rashba spin-orbit interaction.
We derive general spin-charge kinetic equations which show that such spin-orbit interaction,
together with anisotropic Elliott-Yafet spin relaxation, yields significant corrections to
the magnetization-induced dynamics. 
{In particular, we present a consistent treatment of the spin density and spin current contributions
to the equations of motion, inter alia identifying a novel term in the effective force which appears
due to a spin current polarized parallel to the magnetization. This `inverse spin filter' contribution
depends markedly on the parameter which describes the anisotropy in spin relaxation.
To further highlight the physical meaning of the different contributions}, 
the spin pumping configuration of typical experimental setups is analyzed in detail. 
In the two-dimensional limit the build-up of a DC voltage 
is dominated by the spin galvanic (inverse Edelstein) effect.
A measuring scheme that could isolate this contribution is discussed.
\end{abstract}
\maketitle

%%%%%%%%%%%%%%%%%%%%%%%%%%%%%%%%%%%%%%%%%%%%%%%%%%%%%%%%%%%%%%%%%%%%%%%%%%% INTRODUCTION %%%%%%%%%%%%%%%%%%%%%%%%%%%%%%%%%%%%%%%%%%%%%%%%%%%%%%%%%%%%%%%%%%%%%%%
\section{Introduction}
\label{Sec_intro}
The active control of the spin degrees of freedom in a solid state system 
is the central concern of spintronics.\cite{zutic2004} 
The exchange coupling between the magnetization and the spin of charge carriers
is routinely exploited two ways: to generate spin currents and non-equilibrium spin polarizations, 
and to employ such currents and polarizations---generated by other means---to exert a torque on the magnetization.
In this work we are concerned with the first scenario only, though all setups that will be discussed can, and typically are,
used for both purposes.
  
In this context, spin pumping\cite{urban2001,tserkovnyak2002,azevedo2011,czeschka2011} 
and the inverse spin Hall effect (ISHE) \cite{saitoh2006,kimura2007,mosendz2010_2,czeschka2011,obstbaum2014} 
are the tools of choice for generation and detection of electronic spin currents, respectively.
The typical spin pumping setup consists of a magnet\footnote{Magnetic insulators, ferro- or ferrimagnets are used as `spin pumpers' 
in different experiments.  For our purposes the difference between them is actually irrelevant, and we will thus
speak generally about `magnets' or `magnetic materials'.}/normal-metal bilayer.  
The magnetization of the magnetic material is driven such that it performs a conical precession, 
and a spin current perpendicular to the interface (here, along the $z$-direction) builds up, 
${\bf j}_z\sim g^{\uparrow\downarrow}_r {\bf n}\times\dot{\bf n} + g^{\uparrow\downarrow}_i \dot{\bf n}$.
The vector components of ${\bf j}_z$, i.e., $j_z^a$, with $a=x,y,z,$ represent the spin polarization,
${\bf n}$ is the instantaneous magnetization direction, and $g^{\uparrow\downarrow}_r$ ($g^{\uparrow\downarrow}_i$) is the real (imaginary)
part of the spin-mixing conductance $g^{\uparrow\downarrow}$.\cite{tserkovnyak2002} 
Due to the ISHE {in the bulk of the normal metal}, 
this spin current can be detected by measuring the inverse spin Hall voltage {appearing therein}. 
The inverse spin Hall voltage is associated with the spin Hall angle $\theta^\mathrm{sH}$,
which is defined as the ratio of the spin Hall and charge conductivities.
Large spin Hall angles are typically found in transition metals such as Au,\cite{seki2008,obstbaum2014}
Pt\cite{vila2007,kimura2007,liu2011,hahn2013,obstbaum2014} or Ta.\cite{liu2012,hahn2013}
{The same class of setups is also used to study the reciprocal effect, when spin currents generated 
in the normal layer enter the magnetic material and exert a torque on its magnetization.\cite{ando2008,liu2011,liu2012,pesin2012}
The spin-galvanic effect (SGE), \cite{ganichev2002,ganichevreview2011}
which can be related to the ISHE,\cite{raimondi2006,borge2014} represents another channel for spin-to-charge conversion.
It is also referred to as the inverse Edelstein effect,\cite{sanchez2013,shen2014}
and consists in the generation of a charge current perpendicular to the polarization of a nonequilibrium spin density.
Of course, its inverse can as well be used to induce a torque on magnetizations.\cite{avci2014, schulz2015}}

Besides the magnet/normal-metal system just discussed, 
different spin pumping setups are possible.  
For example, spin-charge coupled transport in a Fe/GaAs bilayer can be understood as taking place in
an effective two-dimensional (2D) magnetized electron gas at the Fe/GaAs interface,\cite{gmitra2013, hupfauer2015} 
which can be regarded as a magnet/normal-metal system with the normal metal in the 2D-limit.
Indeed, experimentally realized thin films span the range of thicknesses 
from a few monolayers \cite{shikin2008,rybkin2010} up to tens of nanometers,\cite{valenzuela2006,seki2008,hahn2013}
so that the full three-dimensional (3D) to 2D range is available.  Clearly, the analysis of spin pumping is different
in the 3D or 2D scenario.  In the latter case no spin current can flow perpendicular to the 2D metal, while in-plane
spin currents will be generated by the driving magnetization 
{as soon as in-plane spin-orbit coupling is taken into account, thus leading to in-plane ISHE physics.}  
We will concentrate on the 2D to quasi-2D regime,
in a sense to be made more precise later, and connect our analysis to the one usually performed for 3D systems
in the closing.  Note that this kind of 2D analysis is also relevant for magnet/topological-insulator structures,
which have been recently employed for both spin pumping and reciprocal torque-inducing purposes,\cite{shiomi2014,mellnik2014}
due to the intrinsic 2D nature of the topological surface states.

Typically, spin pumping is most effective as long as the thickness of the film does not exceed 
the spin relaxation length of the normal metal.\cite{tserkovnyak2002,azevedo2011} %{mosendz2010_1,mosendz2010_2}
In thin films, on the other hand, Elliott-Yafet scattering, an important spin-relaxation mechanisms 
in various metals,\cite{zutic2004} should be more effective for in-plane spins than for out-of-plane ones---in
the 2D limit it does not lead to any relaxation of the out-of-plane spins at all.\cite{raimondi2009}
Furthermore, corrections arise in the presence of magnetic textures and intrinsic spin-orbit fields,
and indeed such corrections turn out to be necessary for the physical consistency of the spin dynamics.
These corrections are taken into account, as well as anisotropic spin relaxation.

{While the magnetization of the spin pumping setups mentioned above is homogeneous, 
the situation is even more interesting in the presence of magnetic textures/spin waves,
when complex spin-charge and magnetization dynamics takes place.\cite{tserkovnyak2008_1, tserkovnyak2008_2, vlaminck2008, kajiwara2010}
Hence, we will consider the general situation where the driving is due to a time-dependent magnetic texture,
whose spatial and temporal profile can have any form, and only needs to be smooth 
on the Fermi wavelength and energy scales.}
We will model the thin metallic system as a nearly free electron gas,
and employ an SU(2)-covariant kinetic formulation\cite{gorini2010} 
to compute the effective forces which act on the conduction electrons.
The latter are generated by the interplay of the magnetization dynamics and spin-orbit coupling.
We remark that our kinetic treatment can include finer details of the spin-orbit field, such as those described in
Ref.~\onlinecite{gmitra2013}, where the latter is shown to depend on the direction of the main (static) component of
the magnetization. However, in order to focus on the essentials and avoid overburdening 
the equations, we assume a Rashba-like spin-orbit field only.
Such an effective field is taken to be homogeneous across the whole---not necessarily strictly 2D---sample, 
similar to Refs.~\onlinecite{pesin2012,toelle2014}.  The opposite limit of a bulk metal
with a sharp $\delta$-like Rashba spin-orbit coupling at the interface has also been considered,
\cite{wang2013,haney2013,borge2014,chen2015} and recently discussed in great details.\cite{amin2016a,amin2016b}

The outline of the article is as follows. 
We first (Sec.~\ref{Sec_system}) introduce the system, and connect its model form 
to real-world structures.  In so doing we also clarify the meaning of the important parameters related to the physical energy scales 
of the problem.  In Sec.~\ref{Sec_kinetic} we introduce the model in detail, as well as the transport equations for the charge and spin distribution functions.  The general theoretical results, in particular, the derivation of the generalized effective force acting on the conduction electrons in the presence of spin-orbit coupling, are presented in Sec.~\ref{Sec_SCD}.  
Secs.~\ref{Sec_kinetic} and \ref{Sec_SCD} are technically more involved, and can be
skipped by the reader mostly interested in their specific physical consequences.
An experimentally relevant example is dealt with in Sec.~\ref{Sec_pumping}, 
which analyzes the typical spin pumping configuration.  More precisely, we show that the build-up of a DC electric field 
in a narrow metallic film is mainly due to the SGE and substantially modified by spin relaxation, and suggest that this can be probed by comparing a longitudinal 
and an orthogonal measurement on the same sample.  Here we also comment on the connection between the 2D analysis
of spin pumping, and the established 3D one.  A brief conclusion is given in Sec.~\ref{Sec_conclusions}.
Finally, the appendices show the detailed derivation of the collision integrals and of the generalized spin diffusion equations.

%%%%%%%%%%%%%%%%%%%%%%%%%%%%%%%%%%%%%%%%%%%%%%%%%%%%%%%%%%%%%%%%%%%%%% The system and physical energy scales %%%%%%%%%%%%%%%%%%%%%%%%%%%%%%%%%%%%%%%%%%%%%%%%%%%%

\section{The system and its energy scales}
\label{Sec_system}
\begin{figure}[ht]
  \subfigure[]{
\includegraphics{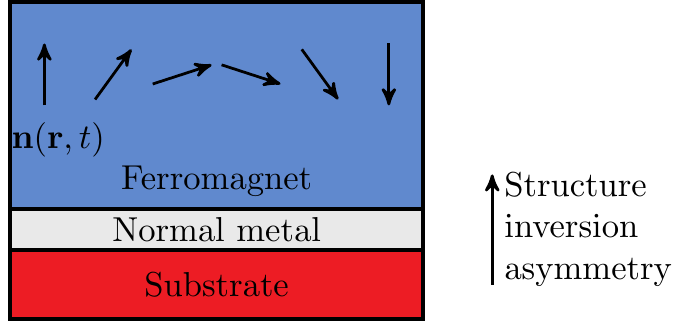}
  }
\quad
 \subfigure[]{
\includegraphics{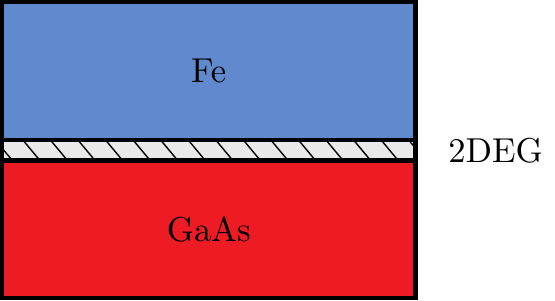}
  }
\caption{{A sketch of the considered structure is shown in (a); ${\bf n}({\bf r},t)$ is the magnetization direction in the magnetic material, 
here illustrated as a N\'eel domain wall. The precise nature of the magnetic layer, e.g., ferro- or ferrimagnetic insulator, is inconsequential 
for our treatment, although it will determine the value of the physical parameters entering the effective Hamiltonian.
The same holds for the normal metal, whose thickness can be anything from a few monolayers (2D) up to tens of
nanometres (3D).  Electrons therein feel an effective Rashba spin-orbit field due to inversion symmetry breaking,
as well as (random) spin-orbit scattering from impurities.  The substrate is a generic structureless insulator, possibly the vacuum.  
Panel (b) shows a possible experimental realization
of a spin pumping setup, where precession of the (here homogeneous) Fe magnetization drives the spin-charge dynamics
of a 2D electron gas formed at the interface with GaAs.} \label{setup}}
\end{figure} 
The system consists of a substrate/normal-metal/magnetic material structure, as sketched in Fig.~\ref{setup}(a),
and is characterized by various energy scales, which will now be introduced.
First of all, the Fermi energy $\epsilon_F$ is assumed to be much larger than any other relevant energy, i.e., we are dealing with a `good metal'.

We then assume a proximity induced magnetization in the metallic film. 
The coupling between the itinerant $s$- and the localized $d$-electrons, i.e., the induced magnetic texture, 
is described within the $s$-$d$ model:
\begin{equation} \label{Eq_sd0}
	H_\mathrm{sd} = \Exc \; {\bn}({\bf r}, t) \cdot \frac{\boldsymbol{\sigma}}{2}\, ,
\end{equation}
where $\boldsymbol{\sigma}=(\sigma^x,\sigma^y,\sigma^z)$ denotes the vector of Pauli matrices, $\Exc$ the ferromagnetic exchange 
band splitting, and ${\bn}$ the magnetization direction.  At first sight this model might appear questionable,
since {we wish to study the dynamics in} the non-magnetic metallic film.  However, it is known that metals like Pt or Pd 
can be magnetized due to the magnetic proximity effect,\cite{antel1999,wilhelm2000,guo2014} and that the exchange energy may be large, 
{i.e., much larger than the disorder broadening, $\Exc\tau/\hbar \gg 1$, with $\tau$ the momentum relaxation time.  
The precise value of $\Exc$
depends on the material properties of the magnetic and non-magnetic layers, and of the interface.}

Furthermore, we assume two types of spin-orbit coupling, a Rashba-like spin-orbit term due to structure inversion asymmetry,
see Fig.~\ref{setup}(a), and extrinsic spin-orbit coupling due to impurities.  
The Rashba spin-orbit coupling Hamiltonian reads
\begin{equation} \label{Eq_Rashba0}
	H_R = -\frac{\alpha}{\hbar} \boldsymbol{\sigma} \times \hat{\bf z} \cdot {\bf p}  \, ,
\end{equation}
with the Rashba parameter $\alpha$, estimated to be between $\num{0.03}$ and $\SI{3}{\electronvolt.\angstrom}$,
depending on the structure and material properties of the system.\cite{ast2007} 
The associated spin-orbit splitting $\Delta_{\rm so}=2\alpha p_F/\hbar$ is taken as small, in the sense that $\Delta_{\rm so}\tau/\hbar\ll1$.  
This condition is often appropriate and, in addition, useful for obtaining physically transparent equations for the spin-charge coupled dynamics.
However, since typical values for the spin-orbit splitting are in the range $10^{-3}\dots 10^{-1}$eV,\cite{shikin2008,rybkin2010,gmitra2013}
this condition is not universally realistic.
Extrinsic spin-orbit coupling with impurities is described by
\begin{equation}
   H_\mathrm{ext} = - \frac{\lambda^2}{4\hbar} \bsigma \times \nabla V({\bf r}) \cdot \bp \, ,
\end{equation}
where $\lambda$ is the effective Compton wavelength, {whose strength is material and impurity-type dependent},
and $V({\bf r})$ is the disorder potential.  
Due to the two types of spin-orbit coupling, both Dyakonov-Perel (DP) and Elliott-Yafet (EY) spin relaxation mechanisms are present. 
The corresponding energies are given by
\begin{eqnarray}
	\frac{\hbar}{\tauDP} = \hbar \left( \frac{2m\alpha}{\hbar^2} \right)^2D\, , \\
	\frac{\hbar}{\tausf} = \frac{\hbar}{\tau} \left( \frac{\lambda p_F}{2\hbar} \right)^4 \, , 
\end{eqnarray}
respectively, with the effective mass $m$, the Fermi momentum $p_F$, and  the diffusion constant $D=v_F^2\tau/d$, 
where $d=2,3$ represents the dimensionality.  
Note that the expression for $\hbar/\tauDP$ follows from the condition $\Delta_{\rm so}\tau/\hbar\ll1$.
Dyakonov-Perel relaxation is intrinsically non-isotropic, since the Rashba term \eqref{Eq_Rashba0}
contains only in-plane momenta, and while its strength depends on the dimensionality of electronic motion through $D$,
its anisotropy does not.  Elliot-Yafet relaxation is, on the other hand, strongly anisotropic only in the 2D limit.
Here, however, `2D' does not refer to the electronic motion, being rather determined by the ratio of the metal
thickness, $t_m$, to the spin relaxation length: Elliott-Yafet is 2D (3D) roughly for $t_m$ small (large) in this sense.  
The transition is modelled by introducing a phenomenological parameter $0 < \zeta < 1$, 
with $\zeta=0 \leftrightarrow 2D,\;\zeta=1\leftrightarrow 3D$ (see Sec.~\ref{Sec_kinetic}).
We focus on the experimentally relevant regime $1 / \tau_\mathrm{DP} \gg 1 / \tausf$.\cite{sanchez2013}
Together with the strong exchange assumption, $\Exc\tau/\hbar\gg1$,\cite{guo2014} this leads to 
the following hierarchy of energy scales:
\be
\label{hierarchy}
\underbrace{\frac{\hbar}{\Exc \tausf}}_{\betasf} \ll \underbrace{\frac{\hbar}{\Exc \tauDP}}_{\betaDP} \ll \frac{\hbar}{\Exc\tau} 
\ll 1 \ll \frac{\hbar}{\Delta_{\rm so}\tau}
\ee
Equation \eqref{hierarchy} defines the spin torque parameters $\betasf$ and $\betaDP$, which will appear repeatedly below.

The magnetization is assumed to be smooth on the Fermi wavelength ($\lambda_F$) scale,
and the frequency of its time-dependence is taken as small compared to the spin-flip rate,
\be
\omega \tau_s / \hbar \ll 1 ,
\ee
applicable for the typical adiabatic pumping regime. In Sec.~\ref{Sec_SCD}, we will in addition consider 
the diffusive regime,
\be
\omega \tau, ql \ll 1 ,
\ee
with $q$ the typical wavevector of the system inhomogeneities, and $l=v_F\tau$ the mean free path.

%%%%%%%%%%%%%%%%%%%%%%%%%%%%%%%%%%%%%%%%%%%%%%%%%%%%%%%%%%%%%%%%%%%%%% THE KINETIC EQUATIONS %%%%%%%%%%%%%%%%%%%%%%%%%%%%%%%%%%%%%%%%%%%%%%%%%%%%%%%%%%%%%%%%%%%%

\section{The kinetic equations}
\label{Sec_kinetic}

In order to describe the transport phenomena in the presence of spin-orbit coupling and a magnetic texture, we use the SU(2) formulation of the Boltzmann-like equation.\cite{gorini2010} The Hamiltonian of the system reads
\begin{equation}
	H = \frac{1}{2m}\left({\bf p} + \boldsymbol{\mathcal{A}}^a \frac{\sigma^a}{2} \right)^2 +e \Phi + \Psi^a({\bf r}, t) \frac{\sigma^a}{2} + V({\bf r}) + H_\mathrm{ext}  \, , \label{hamiltonian_eq}
\end{equation} 
where $\boldsymbol{\mathcal{A}}^a$ is an SU(2) vector potential which describes the intrinsic spin-orbit coupling, $\Phi$ is the electric potential with $e=|e|$, and $\Psi^a$ is an SU(2) scalar potential. Here and throughout the paper, upper (lower) indices will indicate spin (real space) components. A summation over repeated indices is implied.

For the system as discussed in Sec.~\ref{Sec_system}, we have
\begin{eqnarray} 
	H_\mathrm{sd}  \hspace*{0.1cm} &\longleftrightarrow& \hspace*{0.1cm}  \Psi^a({\bf r}, t)  = \Exc \, n^a({\bf r}, t) \, , \label{Eq_sd} \\
	H_R \hspace*{0.1cm} &\longleftrightarrow& \hspace*{0.1cm} \mathcal{A}_y^x=- \mathcal{A}_x^y = \frac{2m\alpha}{\hbar} \, , \label{Eq_Rashba}
\end{eqnarray}
compare Eqs.~(\ref{Eq_sd0}) and (\ref{Eq_Rashba0}).

According to Ref.~\onlinecite{gorini2010} and for $\delta$-correlated (short-range) disorder, the Boltzmann equation for the distribution function $f=f^0+{\bf f}\cdot\boldsymbol{\sigma}$, with $f^0$ (${\bf f}$) the particle (spin) distribution function, reads
\begin{equation}\label{Eq_Boltzmann}
 \tilde{\partial}_t f + \frac{\bp}{m} \cdot \tilde{\nabla}_\br f + \frac{1}{2} \left\lbrace \bcalF , \nabla_\bp f\right\rbrace  = -\frac{1}{\tau} \left( f- \langle f \rangle \right) + I_\mathrm{EY}[f] \, ,
\end{equation}
where $\langle \dots \rangle$ denotes the angular average w.r.t.\ the momentum. The covariant time (spatial) derivative $\tilde{\partial}_t$ ($\tilde{\nabla}_{\bf r}$) and the generalized force $\boldsymbol{\mathcal{F}}$ are given by
\begin{eqnarray}
	\tilde{\partial}_t &=& \partial_t - \frac{i}{\hbar} \left[ \Psi^a \frac{\sigma^a}{2} , \cdot \right] \, , \label{Eq_covtimederiv} \\
	\tilde{\nabla}_{\bf r} &=& \nabla_{\bf r} + \frac{i}{\hbar} \left[  \boldsymbol{\mathcal{A}}^a \frac{\sigma^a}{2} , \cdot \right] \, , \label{Eq_covspacederiv} \\
	\boldsymbol{\mathcal{F}} &=& -e{\bf E} - \left( \boldsymbol{\mathcal{E}}+\frac{{\bf p}}{m} \times \boldsymbol{\mathcal{B}} \right)^a \frac{\sigma^a}{2} \, ,
\end{eqnarray}
and
\begin{eqnarray}
	\mathcal{E}_i^a &=& -\nabla_i \Psi^a - \frac{1}{\hbar} \varepsilon^{abc} \Psi^b \mathcal{A}_i^c \, , \\
	\mathcal{B}_i^a &=& -\frac{1}{2\hbar} \varepsilon_{ijk} \varepsilon^{abc} \mathcal{A}_j^b \mathcal{A}_k^c \, ,
\end{eqnarray}
$\nabla_i$ denoting the $i$-th component of $\nabla_{\bf r}$. The dot within the commutator in Eqs.~(\ref{Eq_covtimederiv}) and (\ref{Eq_covspacederiv}) is a placeholder for the object on which the covariant derivative acts.  Using Eqs.~(\ref{Eq_sd}) and (\ref{Eq_Rashba}) the $i$-th component of the generalized force reads
\begin{equation}
 	\mathcal{F}_i = -eE_i + \Exc  \left(\tilde{\nabla}_i \bn \right)\cdot \frac{\bsigma}{2} - \frac{2m\alpha^2}{\hbar^3} \varepsilon_{ijz} p_j \sigma^z \, .
\end{equation}  
The $i$-th component of the (3D) covariant derivative $\tilde{\nabla}_i$ is defined as
\begin{equation} \label{Eq_CovDev}
	\tilde{\nabla}_i = \nabla_i + [\ba_i]_\times \, ,
\end{equation}
where we have introduced the notation for an antisymmetric matrix $[{\bf v}]_\times$ with its components defined by $([{\bf v}]_\times)^{ab}=-\varepsilon^{abc}v^c$. This definition corresponds to a cross product in the sense that $[{\bf v}]_\times {\bf b}={\bf v} \times {\bf b}$ for an arbitrary vector ${\bf b}$. The vector ${\bf a}_i$ is defined as ${\bf a}_i = 2m\alpha/\hbar^2 (-\delta_{iy},\delta_{ix},0)$. Analogously, from now on we use the `3D covariant time derivative' defined as
\begin{equation}
\tilde{\partial}_t = \partial_t + \frac{\Exc}{\hbar} [\bn]_\times \, .
\end{equation}

\begin{figure}
\includegraphics{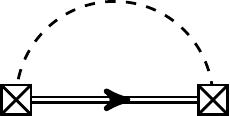}
\caption{Impurity averaged self-energy which determines the Elliott-Yafet collision
operator. The boxed crosses, the dashed line, and the arrowed double line represent spin-orbit coupling,
impurity correlations, and the Green's function in Keldysh space, respectively.} \label{Fig_selfenergy}
\end{figure}

The quantity $I_\mathrm{EY}$ on the r.h.s.\ of Eq.~(\ref{Eq_Boltzmann}) is the Elliott-Yafet collision operator, representing spin-flip processes. It follows from the impurity averaged self-energy as depicted in Fig.~\ref{Fig_selfenergy}, and is substantially
modified in the presence of intrinsic spin-orbit coupling and magnetic textures.  The corrections are obtained via
a first-order SU(2) shift (see App.~\ref{App_EY}), which yields the following generalized collision integral:
\begin{equation}
I_\mathrm{EY} = I^0_\mathrm{EY} + \delta I^\Psi_\mathrm{EY} + \delta I^{\bcalA}_\mathrm{EY} \ ,
\end{equation}
where
\begin{widetext}
\begin{eqnarray}
I^0_\mathrm{EY} &=& -\frac{1}{2 N_0 \tau} \left( \frac{\lambda p}{2\hbar} \right)^4 \!\!\int\!\! d\bp' \delta(\epsilon_\bp-\epsilon_{\bp'}) \left[\Gamma \left({\bf f}_{\bp} + {\bf f}_{\bp'}  \right)\right] \cdot \bsigma \, \label{Eq_EY} ,\\
\delta I^\Psi_\mathrm{EY} &=&  \frac{ m p^2}{N_0\tau}  \left( \frac{\lambda}{2\hbar} \right)^4  \!\!\int\!\! d\bp' \delta(\epsilon_\bp-\epsilon_{\bp'})  (\Gamma \boldsymbol{\Psi}) \cdot \left[ f^0_\bp \bsigma + {\bf f}_\bp - \left(1 + \epsilon_{\bp'} \partial_{\epsilon_{\bp'}}\right)\Big(  f^0_{\bp'}  \bsigma -   {\bf f}_{\bp'}  \Big) \right]  , \label{Eq_EY1} \\
\delta I^{\bcalA}_\mathrm{EY} &=& \frac{1}{N_0 \tau} \left( \frac{\lambda}{2\hbar} \right)^4  \!\!\int\!\! d\bp' \delta(\epsilon_\bp-\epsilon_{\bp'}) \bcalA^a \cdot {\bf L}_{\bp,\bp'}  \left[ \sigma^a \left( f^0_\bp - f^0_{\bp'} \right) + \left( f^a_\bp + f^a_{\bp'} \right) \right]  , \label{Eq_EY2}
\end{eqnarray}
\end{widetext}
with $\boldsymbol{\Psi} = \Exc \, \bn$,
$d \bp' \equiv d^dp'/(2\pi \hbar)^d$, ${\bf L}_{\bp,\bp'} = (p'^2 + \bp \cdot \bp')\bp - (p^2 + \bp \cdot \bp') \bp'$, the density of states per volume and spin $N_0$, and $\Gamma=\mathrm{diag}(1,1,\zeta)$. The latter takes into account the anisotropy of spin-flip processes, as discussed above,
hence depends on the thickness $t_m$ of the normal metal. 
Clearly $0 \leq \zeta \leq 1$, with $\zeta = 0$ representing the limit that the normal metal is a 2D gas, i.e., for small $t_m$, whereas one may assume $\zeta=1$ when $t_m>\lsf$, where $\lsf$ is the {spin relaxation} 
length of the normal metal. We emphasize that Eqs.~\eqref{Eq_EY1} and \eqref{Eq_EY2} are valid for arbitrary spin-orbit fields,
not only Rashba coupling, and magnetic textures.

The above expressions, Eqs.~(\ref{Eq_EY1}) and (\ref{Eq_EY2}), are `first-order' in the SU(2) fields, provided we take the spin distribution function $f$ to be `zero-order'. However, as discussed in the next section in connection with Eq.~(\ref{Eq_distrfunction}), ${\bf f}$ contains a local-equilibrium part, ${\bf f}_\mathrm{eq}$, which formally is also `first-order'. Thus, in order to treat relaxation due to the EY collision operator consistently, it becomes necessary to include also a specific second-order correction in the SU(2) shift, which is given by
\begin{equation}
\delta I_\mathrm{EY}^{\bcalA,\Psi} = - \frac{1}{4\tau} \left( \frac{\lambda}{2\hbar}\right)^4 \boldsymbol{\Psi} \cdot \bcalA_i p_i p^2 \partial_{\epsilon_\bp} \left( \epsilon_\bp \partial_{\epsilon_\bp} f^0_\mathrm{eq} \right) \, ,  \label{Eq_EY3}
\end{equation}
where $f^0_\mathrm{eq}$ is the Fermi function.
As a consequence, only the non-equilibrium part of the spin density, $\delta \bs$, will enter the effective force, Eq.~(\ref{Eq_Fs}).
An even more detailed investigation of the EY collision operator is well underway.\cite{gorini2016}

%%%%%%%%%%%%%%%%%%%%%%%%%%%%%%%%%%%%%%%%%%%%%%%%%%%%%%%%%%%%%%%%%% SPIN-CHARGE COUPLED DYNAMICS %%%%%%%%%%%%%%%%%%%%%%%%%%%%%%%%%%%%%%%%%%%%%%%%%%%%%%%%%%%%%%%

\section{spin-charge coupled dynamics}
\label{Sec_SCD}
Here we present the coupled equations for the electron density, the electron current, the spin density, and the spin current,
respectively defined as follows:\footnote{Lower indices refer to the spatial component, whereas upper indices represent the polarization. When the spin current is written in boldface this means a vector consisting of the three components which are not marked as index.}
\begin{eqnarray}
	n &=& 2\int \!\! d \bp \, f^0 \, , \\
	j_i &=& 2\int \!\! d \bp \, \frac{p_i}{m} f^0  \, , \\
	\bs &=& \int \!\! d \bp \, {\bf f} \, , \\
	j_i^a &=& \int \!\! d \bp \, \frac{p_i}{m} f^a \, .
\end{eqnarray}
{We focus first on the spin sector (\ref{Subsec_spin}), and second on the charge sector (\ref{Subsec_charge}), 
and third discuss the interpretation of the different contributions to the effective force (\ref{Subsec_discussion}).}

\subsection{Spin sector}
\label{Subsec_spin}
In order to study the spin sector, we multiply the Boltzmann equation with the Pauli vector $\bsigma$ and perform the trace. Before doing so, it is convenient to split the spin distribution function ${\bf f}$ as 
\begin{equation} \label{Eq_distrfunction}
{\bf f} = {\bf f}_\mathrm{eq} + \delta {\bf f}
\end{equation}
with ${\bf f}_\mathrm{eq} = \left(-\partial_{\epsilon_\bp} f^0_\mathrm{eq}\right) (\Exc/2)\bn$, where $f^0_\mathrm{eq}$ is the Fermi function. This is motivated by the form of the spin density
\begin{equation}
{\bs} = {\bs}_\mathrm{eq} + \delta {\bs} \, ,
\end{equation} 
where ${\bs}_\mathrm{eq}=(N_0\Exc/2)\bn$ is the equilibrium part of $\bs$ which adiabatically follows the magnetization.  
The dynamics of the itinerant electrons, which is typically much faster than the magnetization dynamics, 
leads to the nonequilibrium contribution $\delta \bs$.

We trace over the spin sector and obtain the following $3 \times 3$ matrix equation:
\begin{equation} \label{Eq_spinsector}
\mathbb{M} \delta {\bf f} = \mathbb{N} \left\langle{ \delta \bf f}\right\rangle + {\bf S} \, ,
\end{equation}
with
\begin{eqnarray}
\mathbb{M} &=&   1+\frac{\tau}{2\tausf} \Gamma +\tau \tilde{\partial}_t  + \tau \frac{p_i}{m}\tilde{\nabla}_i \, , \\ 
\mathbb{N} &=&  1 - \frac{\tau}{2\tausf} \Gamma \, , \\
{\bf S} &=& \left(\partial_{\epsilon_\bp} f^0_\mathrm{eq}\right) \frac{\tau \Exc}{2} \dot{\bn} \, .
\end{eqnarray}
Note that we have neglected small deviations of $f^0$ from its angular average, $f^0 \simeq \langle f^0 \rangle $, since these are at least first order in the electric field ${\bf E}$ or the magnetic texture, i.e., $\nabla_i \bn$ or $\dot{\bn}$. Furthermore we assume $\langle f^0 \rangle \simeq f^0_\mathrm{eq}$.

By an integration of the spin sector over the momentum we obtain
\begin{equation} \label{Eq_contEq0}
\tilde{\partial}_t \delta \bs + \tilde{\nabla}_i{\bf j}_i = -\frac{1}{\tausf} \Gamma \delta \bs - \frac{N_0 \Exc}{2} \dot{\bn} \, .
\end{equation}
Next, we consider the quasiadiabatic limit, $\tausf \partial_t \delta \bs \ll \delta \bs$, as well as $\tausf \partial_t \delta \bs \ll \zeta \delta \bs$.  We are then able to solve for the nonequilibrium spin density:
\begin{equation}\label{Eq_ssplit}
\delta \bs = \left(\delta\bs\right)_{\bn}  + (\delta\bs)_{\bj_s} \, ,
\end{equation}
with 
\begin{equation} \label{Eq_s0}
\left(\delta\bs\right)_{\bn} =  -\frac{N_0 \Exc\tausf}{2} \left(\Gamma  + \betasf^{-1} [\bn]_\times \right)^{-1} \dot{\bn} 
\end{equation}
the part of $\delta \bs$ which is associated directly with the magnetization, and with
\begin{equation} \label{Eq_ssc}
(\delta\bs)_{\bj_s} = - \tausf \left(\Gamma  + \betasf^{-1} [\bn]_\times \right)^{-1}  \tilde{\nabla}_i  \bj_i
\end{equation}
the part which is associated with the spin current. In general the spin current itself depends on the spin density,
which has to be kept in mind when solving for the spin density from Eq.~(\ref{Eq_contEq0}). The split in Eq.~(\ref{Eq_ssplit}) 
is found to be technically convenient. 

In the following, we shall calculate the spin current in the diffusive regime. Our approach is to rewrite the matrix $\mathbb{M}$ in 
Eq.~(\ref{Eq_spinsector}) as follows:
\begin{equation}
\mathbb{M} = (1+\xi) M \, ,
\end{equation}
where
\begin{equation} \label{Eq_M}
M =  \left(1+\frac{\Exc\tau}{\hbar}[{\bf n}]_\times\right) \, ,
\end{equation}
and
\begin{equation}
\xi = \left(\frac{\tau}{2\tausf} \Gamma +\tau\partial_t  + \tau \frac{p_i}{m}\tilde{\nabla}_i\right) M^{-1} .
\end{equation}
In the diffusive regime we can approximate $(1+\xi)^{-1} \simeq 1 - \xi$, and rewrite Eq.~(\ref{Eq_spinsector}) as
\begin{equation}
 \delta {\bf f} = M^{-1} (1-\xi) \Big(  \mathbb{N} \left\langle{ \delta \bf f}\right\rangle + {\bf S}  \Big) \, .
\end{equation}
By multiplying the latter equation with $p_i/m$ and integrating over the momentum, we obtain the spin current:
\begin{equation} \label{Eq_spincurrent}
\bj_i = \left( \bj_i \right)_{\bn} + \left( \bj_i \right)_{\bs} \, ,
\end{equation}
with 
\begin{equation} \label{Eq_sc_n}
\left(\bj_i\right)_{\bn} = \frac{\tau}{2} D N_0 \Exc  M^{-1} \tilde{\nabla}_i M^{-1} \, \dot{\bn}
\end{equation}
the part arising directly from the magnetization, and
\begin{equation} \label{Eq_sc_deltas}
\left(\bj_i\right)_{\bs} =  -D M^{-1} \tilde{\nabla}_i M^{-1} \delta\bs
\end{equation}
the part of the spin current which has its source in the spin density.

\subsection{Charge sector}
\label{Subsec_charge}
For the charge sector, we trace over the Boltzmann equation (\ref{Eq_Boltzmann}), multiply with $p_i$, and integrate over the momentum, with the following result:
\begin{equation} \label{Eq_charge}
	\left(1+\tau\partial_t\right) j_i + D \nabla_i n = - n \mu E_i + \frac{\tau N_0 \Exc}{m} F_i \, ,
\end{equation}
where $\mu = e\tau/m$ is the electron mobility, and $n \mu = \sigma_D /e$ with $\sigma_D$ the Drude conductivity. The effective force $F_i$ combines the contributions of the nonequilibrium part of the spin density and the spin current. 

\begin{figure}[bt]
\vspace*{-0.3cm}\hspace*{-1.6mm}
\includegraphics{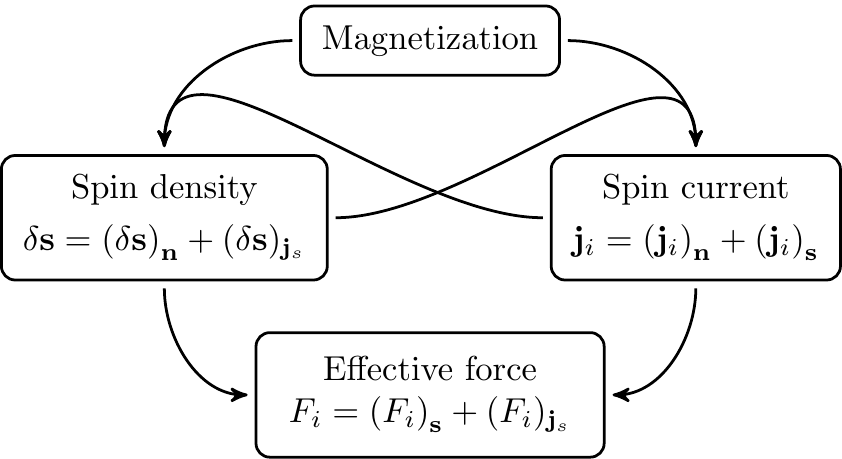} 
\caption{Scheme of the various contributions to the effective force.} \label{Fig_force}
\end{figure}
A scheme of the various contributions to the effective force is depicted in Fig.~\ref{Fig_force}. We split $F_i = \left(F_i\right)_{\bs} + (F_i)_{\bj_s}$ 
into two terms which represent the contribution of the spin density, $(F_i)_{\bs}$, and a term representing the direct contribution of the spin current, 
$(F_i)_{\bj_s}$. According to this split it is clear that $(F_i)_{\bs}$ is associated with the SGE, and $(F_i)_{\bj_s}$ with the ISHE. The two 
contributions to the effective force explicitly read
\begin{eqnarray}
(F_i)_{\bs} &=& \frac{1}{N_0} \left[ \left( \tilde{\nabla}_i \bn\right) \cdot \delta \bs + \frac{2m\alpha}{\hbar^2} \betasf (\bz \times \delta \bs)_i\right] \, , 
\label{Eq_Fs}\\
(F_i)_{\bj_s} &=& \frac{1}{D N_0} \left[\betaDP \left( \bj^z \times \bz \right)_i + \frac{\tau}{\tausf} \bnz \cdot \bj_i \right] \, , 
\label{Eq_Fj}
\end{eqnarray}
with $\bnz = \Gamma \bn$. With respect to the second term on the r.h.s.\ of Eq.~(\ref{Eq_Fs}), compare the discussion in connection with Eq.~(\ref{Eq_EY3}).

We further divide $(F_i)_{\bs}$ into contributions arising from $(\delta\bs)_\bn$ and $(\delta\bs)_{\bj_s}$:
\begin{equation}
 (F_i)_{\bs}=(F_i)_{\bs, \bn} + (F_i)_{\bs, {\bj_s}} \, , \label{Eq_Fi_spindensity}
\end{equation}
where $(F_i)_{\bs,\bn}$ and $(F_i)_{\bs,{\bj_s}}$ have the same form as $(F_i)_{\bs}$ in Eq.~(\ref{Eq_Fs}), but with $\delta\bs$ being replaced by $(\delta\bs)_\bn$ and $(\delta\bs)_{\bj_s}$, respectively. Analogously, we define 
\begin{equation}
(F_i)_{\bj_s} = (F_i)_{{\bj_s},\bn}  + (F_i)_{{\bj_s},\bs}  \label{Eq_Fi_spincurrent}
\end{equation}
with $(F_i)_{{\bj_s},\bn}$ and $(F_i)_{{\bj_s},\bs}$ of the same form as $(F_i)_{\bj_s}$ in Eq.~(\ref{Eq_Fj}), but with $\bj_i$ being replaced by $(\bj_i)_\bn$ and $(\bj_i)_{\bs}$, respectively.
{The idea behind this separation is to express the effective force in terms of the drive, $\dot\bn$, and the spin current, the subscripts 
indicating the respective origins.}

\subsubsection{The contribution $(F_i)_{\bs, \bn}$} 
The spin density related directly to the dynamical magnetization, see Eq.~(\ref{Eq_s0}), is explicitly given by
\begin{equation} \label{Eq_sn}
 \left(\delta\bs\right)_{\bn} = \frac{\hbar N_0}{2} \frac{1}{\bnz^2} \bigg[ \bnz \times \dot{\bn} - \betasf \zeta \Gamma^{-1} \dot{\bn} \bigg] \, .
\end{equation}  
 We insert Eq.~(\ref{Eq_sn}) into Eq.~(\ref{Eq_Fs}) and find
\begin{eqnarray} 
 (F_i)_{\bs,\bn} &=&\frac{\hbar}{2} \frac{1}{\bnz^2} \Bigg\lbrace  \left(\nabla_i \bn \right) \cdot \left( \bnz \times \dot{\bn} - \betasf\zeta {\Gamma}^{-1} \dot{\bn}  \right) \nonumber \\
 &&{}+ \frac{2m\alpha}{\hbar^2} \big[ \left(\bn \cdot \bnz \right) \bz \times \dot{\bn} \nonumber \\
 &&{}+ \betasf \bz \times \bm{(} \bnz \times \dot{\bn} + \zeta \bn \times \Gamma^{-1} \dot{\bn}  \bm{)} \big]_i \Bigg\rbrace \, . \label{Eq_F0_full}
\end{eqnarray}
Assuming $\zeta=1$ (i.e., $t_m \gg \lsf$) the last equation reduces to Eq.~(11) in Ref.~\onlinecite{knoester2014}; see also 
Refs.~\onlinecite{tserkovnyak2008_2,stern1992,barnes2007,duine2008,tatara2013,yamane2013}. Recall that $\zeta$ describes the anisotropy of spin-flip 
relaxation and that $\zeta=1$ corresponds to the isotropic case. As far as we know, this is the first time that such anisotropy ($\zeta < 1$) is 
explicitly taken into account. However, experiments on thin films typically deal with samples on the scale of a few nanometres, hence it is appropriate 
to include this effect.

\subsubsection{The contributions $(F_i)_{\bs, {\bj_s}}$ and $(F_i)_{{\bj_s},\bn}$\\
               (homogeneous case)}
For the sake of simplicity and since we are mostly interested in the competition between the SGE and the (in-plane) ISHE, we shall focus here on the Rashba contribution, i.e., we consider $\tilde{\nabla}_i \approx [\ba_i]_\times$ in Eqs.~(\ref{Eq_ssc}) and (\ref{Eq_Fs}), which corresponds to a spatially homogeneous situation. Neglecting terms $\sim\betasf^2$ and smaller 
for the spin density dependent contribution, we find
\begin{eqnarray}
(F_i)_{\bs,{\bj_s}} &=& \frac{1}{D N_0} \frac{1}{\bnz^2} \betaDP \bigg\lbrace \left( \bn \cdot \bnz \right) \big[ \bz \times \left( \bj^z \right)_\bn\big]_i  \nonumber \\ 
+ & 2 & \sum\limits_{a=x,y}\!\!\big[ n_z \left( j^a_a \right)_\bn \!-\! n_a \left( j^z_a \right)_\bn \big] \!  \left( \bz \times \bn \right)_i \!\! \bigg\rbrace . \label{Eq_Fsjs}
\end{eqnarray}
Adding the contribution by $(\bj_i)_\bn$ [Eq.~(\ref{Eq_Fj}) with $\bj_i \rightarrow (\bj_i)_\bn$] we obtain
\begin{eqnarray} 
&(F_i)_{\bs,{\bj_s}}& + (F_i)_{{\bj_s},\bn} \nonumber \\
&&{}=  \frac{1}{DN_0} \Bigg\lbrace  
\frac{\betaDP\zeta(1-\zeta)n_z^2}{\bnz^2}\left[ \bz \times  \left( \bj^z \right)_\bn \right]_i \nonumber \\
&&{}\phantom{=}+ 2\betaDP\sum\limits_{a=x,y}\big[ n_z \left( j^a_a \right)_\bn - n_a \left( j^z_a \right)_\bn \big] \left( \bz \times \bn \right)_i \Bigg\rbrace \nonumber \\
&&{}\phantom{=}+ \frac{\tau}{\tausf} \bnz \cdot \left(\bj_i\right)_\bn \, . \label{Eq_Fscj}
\end{eqnarray}
Note that $(F_i)_{{\bj_s},\bn}$ features a direct contribution due to the driving source, i.e., $\dot{\bn}$, whereas $(F_i)_{{\bj_s},\bs}$ contributes more indirectly through $\delta\bs$. Apparently Eq.~(\ref{Eq_Fscj}) yields a non-trivial interplay between the two `origins' (spin density versus spin current, cf.\ Eqs.~(\ref{Eq_Fi_spindensity}) and (\ref{Eq_Fi_spincurrent})) since the first term on the r.h.s.\ of Eq.~(\ref{Eq_Fj}) is cancelled to some extent by the first term on the r.h.s.\ of Eq.~(\ref{Eq_Fsjs}). This 
demonstrates once more that the interplay between SGE and ISHE is non-trivial. 

\subsection{Discussion: effective forces}
\label{Subsec_discussion}
{We emphasize that the above expressions, Eqs.~(\ref{Eq_Fs}) and (\ref{Eq_Fj}), obtained by properly integrating the kinetic equation, 
are of general validity. Before going into the details of the spin pumping configuration (Sec.\ \ref{Sec_pumping}),
we comment on the relation of these equations to previous results. First, 
neglecting the Rashba contribution in the first term on the r.h.s.\ of Eq.~(\ref{Eq_Fs}), i.e., $\tilde{\nabla}_i \bn \to {\nabla}_i \bn$, considering
the isotropic case ($\zeta = 1$), and taking into account that
\begin{equation}
\delta\bs = \frac{\hbar N_0}{2} [ \bn \times \dot{\bn} - \betasf \dot{\bn} ]
\end{equation}
in this limit, this force term agrees with the result given in Ref.~\onlinecite{tserkovnyak2008_2}. Second, including the Rashba contribution but neglecting the spin
current contribution to $\delta\bs$, cf.\ Eq.~(\ref{Eq_ssc}), and again for $\zeta = 1$, Eq.~(\ref{Eq_Fs}) reduces to Eq.~(11) in Ref.~\onlinecite{knoester2014}.
Third, the first term on the r.h.s.\ of Eq.~(\ref{Eq_Fj}), which is due to Rashba spin-orbit coupling, is related to the ISHE: a charge current in the $x$-$y$-plane is
generated by a spin current (in that plane) polarized in $z$-direction, with the charge current direction being perpendicular to the spin current direction,
$\sim \bj^z \times \bz$, cf.\ Ref.~\onlinecite{saitoh2006}.
}

{Note, however, that the various terms are not independent from each other. In particular,
spin density and spin current are closely related, as already pointed out in Ref.~\onlinecite{raimondi2012}, due to the interplay of Rashba 
coupling and EY relaxation. This is apparent from Eq.~(\ref{Eq_contEq0}), which for time-independent and spatially homogeneous situations reads:
\begin{equation}
\label{Eq_contEq0_homo}
\frac{\Exc}{\hbar} \bn \times \delta \bs + \ba_i \times {\bf j}_i = -\frac{1}{\tausf} \Gamma \delta \bs \, .
\end{equation}
Taking this equation into account, it is possible to relate the total effective force derived here to the one discussed in Ref.~\onlinecite{shen2014} 
(see Eq.~(12) therein). First, consider the limit $\Exc \to 0$, which leaves only the second term in (\ref{Eq_Fs}) and the first term in (\ref{Eq_Fj}).
The latter is readily identified with the `Hall-like' force;\cite{shen2014} however, using Eq.~(\ref{Eq_contEq0_homo}) we find that the new term,
Eq.~(\ref{Eq_Fs}), which results from the EY collision operator, Eq.~(\ref{Eq_EY2}), gives
an identical contribution, thus our result appears to be larger by a factor of two. Further differences become apparent for finite $\Exc$.
}

{Last but not least, the second term on the r.h.s.\ of Eq.~(\ref{Eq_Fj}) is denoted `inverse spin filter' term, as it
describes a force arising from a spin current which is polarized parallel the magnetization (roughly speaking), its strength being $\sim \tausf^{-1}$.
To the best of our knowledge, such a term has not been explicitly considered before. However, it can be related to the anomalous Hall
effect: imagine that an electric field in $x$-direction creates a spin current via the spin Hall effect (in $y$-direction, polarized in $z$-direction).
This spin current leads via the inverse spin filter term to a charge current in $y$-direction. Note that a non-zero $\zeta$ is required for this argumentation to be
valid. In this context, see also the discussions given in Refs.~\onlinecite{schwab2010} and \onlinecite{guo2014}.
}

%%%%%%%%%%%%%%%%%%%%%%%%%%%%%%%%%%%%%%%%%%%%%%%%%%%%%%%%%% SPIN PUMPING AND ISHE %%%%%%%%%%%%%%%%%%%%%%%%%%%%%%%%%%%%%%%%%%%%%%%%%%%%%%%%%%%%%%%%%%%%

\section{Spin pumping configuration}
\label{Sec_pumping}
\begin{figure}[ht]
    \subfigure[]{
    \includegraphics{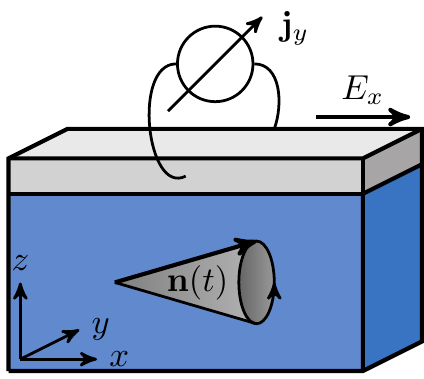}
    }
	   %\hfill
    \quad
    \subfigure[]{\includegraphics[width=0.3\columnwidth]{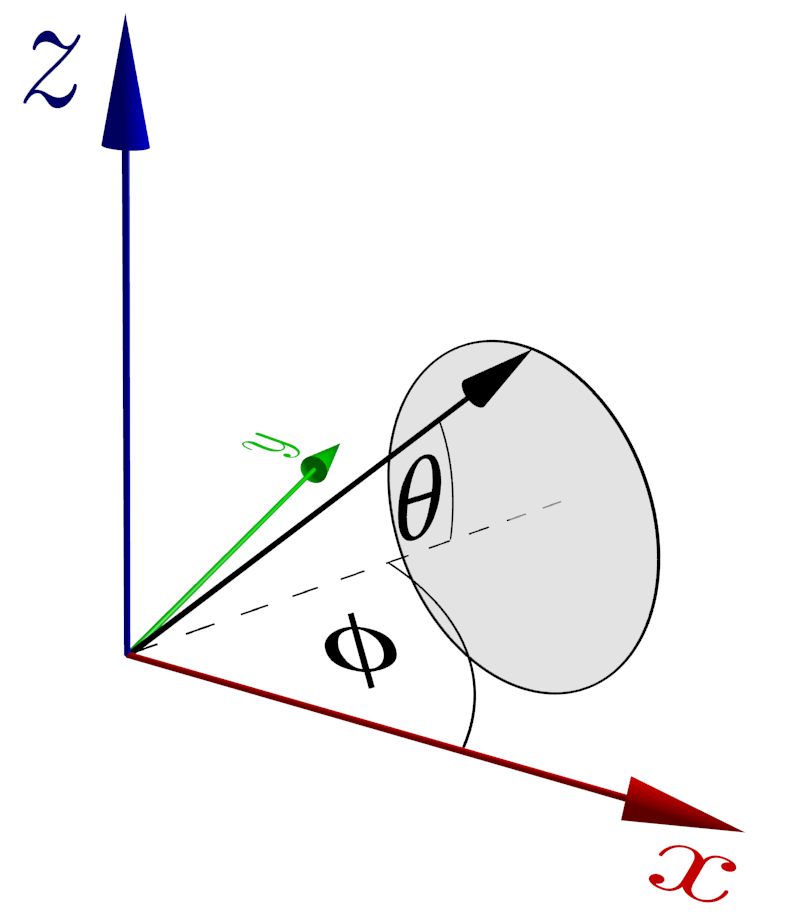} }%
\caption{ (a) The studied configuration, i.e., a metallic film on top of a magnetic material (shown in blue). (b) Sketch of the conical precession of the magnetization, defining the relevant angles $\theta$ and $\phi$. \label{Cone}}
\end{figure} 
 
In this section we consider the magnetization dynamics to be fixed, namely as a precession with a cone angle $\theta$ 
and angular frequency $\omega$ about an axis fixed by an external static and homogeneous magnetic field.
\footnote{Experimentally, the setup employed to excite the magnetization dynamics has to be carefully chosen,
see the discussion in Ref.~[\onlinecite{obstbaum2014}].} 
The magnetization direction is parametrized as follows:
\begin{equation} \label{Eq_bn}
 	{\bf n}(t) = R_\phi \begin{pmatrix}
 	n_0 \\ \delta n_y(t) \\ \delta n_z(t)
 	\end{pmatrix} = R_\phi \begin{pmatrix}
 	\cos\theta \\ \sin\theta\cos\omega t \\ \sin\theta\sin\omega t
 	\end{pmatrix}  \, ,
\end{equation} 
with $R_\phi$ a rotation matrix around the $z$-axis,
\begin{equation}
	R_\phi = \begin{pmatrix}
	\cos \phi & -\sin \phi & 0 \\
	\sin \phi & \cos \phi & 0 \\
	0 & 0 & 1 \\
	\end{pmatrix} \, ,
\end{equation}
where $\phi$ is the angle between the $x$-axis and the cone axis, see Fig.~\ref{Cone}.

Furthermore, we assume open circuit conditions in $x$- and $y$-direction, in order to determine the electric field along these directions. Since the magnetization is homogeneous we expect the particle current to be homogeneous as well, and due to the open circuit condition we have $j_{x,y}=0$ in the whole sample. From Eq.~(\ref{Eq_charge}) we find
\begin{equation}
	\sigma_D E_{x,y} = \frac{e \tau N_0 \Exc}{m} F_{x,y} \, . \label{Eq_Ex}
\end{equation}
According to Ref.~\onlinecite{raimondi2012} the spin Hall conductivity can be expressed 
as\footnote{{Note that in this work we do not consider the side-jump and skew-scattering
contributions. Hence, from the first part of Eq.~(36) in Ref.~\onlinecite{raimondi2012} we obtain for $\tausf \gg \tauDP$, cf.\ Eq.~\eqref{hierarchy},
$\sigma^\mathrm{sH} = (\tauDP/\tausf)\sigma^\mathrm{sH}_\mathrm{int}$ which, together with $\sigma^\mathrm{sH}_\mathrm{int} = e\tau/4\pi\tauDP$,
leads to the result given.}} 
$\sigma^\mathrm{sH} = e\hbar\tau N_0/2m\tausf$ (for $\tausf \gg \tauDP$), hence we may rewrite Eq.~(\ref{Eq_Ex}) as
\begin{equation}\label{Eq_Ex2}
eE_{x,y} = \frac{2\theta^\mathrm{sH}}{\betasf} F_{x,y} \, ,
\end{equation}
where $\theta^\mathrm{sH}=e\sigma^\mathrm{sH}/\sigma_D$ is the spin Hall angle. 
{In order of magnitude, $\theta^\mathrm{sH}\sim \hbar/\epsilon_{F}\tausf$.}
Our focus in the following is on the DC contribution to the electric field, thus we will average Eq.~(\ref{Eq_Ex2}) with respect to time.

Let us now explicitly consider the $x$-component of the effective force. According to Eqs.~(\ref{Eq_F0_full}) and (\ref{Eq_Fscj}), as well as
Eq.~(\ref{Eq_Fj}) with $\bj_i \rightarrow (\bj_i)_\bs$, and it is the sum of the following three contributions:
\begin{eqnarray}
(F_x)_{\bs,\bn} &=& -\frac{m\alpha}{\hbar}\bigg[ \dot{n}_y + \betasf \left(1+\zeta\right) \left( \bn \times \dot{\bn} \right)_y \bigg]  \, , \label{Eq_F_xsn} \\
(F_x)_{\bs,{\bj_s}} + && \!\!\!\!\!\! (F_x)_{{\bj_s},\bn}  \nonumber \\
& =&  - \frac{\betaDP}{DN_0} \bigg\lbrace  \zeta(1-\zeta)n_z^2 \left( j_y^z \right)_\bn \nonumber  \\ 
&{}& + 2 n_y \sum\limits_{a=x,y} \left[n_z\left( j_a^a \right)_\bn - n_a \left( j_a^z \right)_\bn \right] \bigg\rbrace  \nonumber \\
&{}& + \frac{1}{DN_0} \frac{\tau}{\tausf} \bnz \cdot \left(\bj_x \right)_\bn  , \label{Eq_F_xsc} \\
(F_x)_{{\bj_s},\bs} &=& \frac{1}{DN_0} \left[ \betaDP  \left( j_y^z \right)_{\bs} + \frac{\tau}{\tausf} \bnz \cdot \left(\bj_x\right)_{\bs} \right] \, , \label{Eq_F_xjs}
\end{eqnarray} 
where, in order to derive Eqs.~\eqref{Eq_F_xsn} and \eqref{Eq_F_xsc}, we approximated $\bnz^2 \simeq 1$ and $\bnz \cdot \bn \simeq 1$ 
since the cone angle $\theta$ is usually small.\cite{mosendz2010_2}
For this reason, we shall also allow only terms up to $\sin^2\theta$ when performing the time average, hence we neglect terms of order $n_z^2$ since the time average would lead to $\sim \sin^4\theta$ terms. We realize that the first term on the r.h.s.\ of Eq.~(\ref{Eq_F_xsc}), which has its origin in the interplay of the spin density and the spin current, is negligible.

We rewrite Eqs.~(\ref{Eq_F_xsn})--(\ref{Eq_F_xjs}) in order to elucidate the different effects:
\begin{eqnarray}
F_x^{(A)} &=& \frac{\betaDP}{DN_0} \left( j_y^z \right)_{\bs} \, , \label{Eq_F_ISHEx}\\
F_x^{(B)} &=& -\frac{m\alpha}{\hbar}\bigg[ \dot{n}_y + \betasf \left(1+\zeta\right) \left( \bn \times \dot{\bn} \right)_y \bigg]  \nonumber \\
&{}& - \frac{2\betaDP}{DN_0} n_y \sum\limits_{a=x,y} \left[ n_z \left( j^a_a \right)_\bn - n_a \left( j_a^z\right)_\bn \right]  \, , \label{Eq_F_SGEx} \\
F_x^{(C)} &=& \frac{1}{DN_0} \frac{\tau}{\tausf} \bnz \cdot \bj_x \, , \label{Eq_F_SFx}
\end{eqnarray}
where $F_x^{(A)}$ can be related to the ISHE, and $F_x^{(B)}$ to the SGE. The last term, $F_x^{(C)}$, describes the build-up of an effective force (or electric field) 
due to the spin current polarized parallel to the magnetization,\footnote{More precisely, parallel to $\bnz$.} and is denoted inverse spin filter term, as discussed in
Sec.~\ref{Subsec_charge}.

Analogously, we obtain for the $y$-component:
\begin{eqnarray}
F_y^{(A)} &=& - \frac{\betaDP}{DN_0} \left( j_x^z \right)_{\bs} \, , \label{Eq_F_ISHEy}\\
F_y^{(B)} &=& \frac{m\alpha}{\hbar}\bigg[ \dot{n}_x + \betasf \left(1+\zeta\right) \left( \bn \times \dot{\bn} \right)_x \bigg]  \nonumber \\
&{}& + \frac{2\betaDP}{DN_0} n_x \sum\limits_{a=x,y} \left[ n_z \left( j^a_a \right)_\bn - n_a \left( j_a^z\right)_\bn \right]  \, , \label{Eq_F_SGEy} \\
F_y^{(C)} &=& \frac{1}{DN_0} \frac{\tau}{\tausf} \bnz \cdot \bj_y \, . \label{Eq_F_SFy}
\end{eqnarray}

In the following subsections, we consider a narrow wire (see Fig.~\ref{Fig_Fields}) and the electric field that will be measured in a longitudinal and an orthogonal measurement. We assume that the wire is `narrow' such that the width of the wire is smaller than the spin diffusion length $\lsf = \sqrt{D\tausf}$. For such a configuration, the spin current contribution polarized parallel to the magnetization and flowing parallel to the narrow edge vanishes, see App.~\ref{App_Narrow}.

\subsection{Longitudinal measurement}
Let us consider a narrow wire as depicted in Fig.~\ref{Fig_Fields}. For such a sample we find a homogeneous spin current flowing in $x$-direction which, in particular, has a contribution polarized along $\bn$, giving rise to the force in Eq.~(\ref{Eq_F_SFx}) when performing a longitudinal measurement. We perform the time average of Eqs.~(\ref{Eq_F_ISHEx})--(\ref{Eq_F_SFx}), insert the results into Eq.~(\ref{Eq_Ex2}), and obtain the following DC electric fields:
\begin{eqnarray}
\left\langle E_x^{(A)} \right\rangle_t  &\sim & \betaDP  \theta^\mathrm{sH} \frac{F_\alpha}{e} \ll \theta^\mathrm{sH} \frac{F_\alpha}{e} \, , \\
\left\langle E_x^{(B)} \right\rangle_t  &=& -2 \theta^\mathrm{sH} \frac{F_\alpha}{e} (1+ \zeta)  \sin\phi  \sin^2\theta    \label{Eq_Ex_SGE}  \, , \\
\left\langle E_x^{(C)} \right\rangle_t  &=& - \theta^\mathrm{sH} \frac{F_\alpha}{e} (1-\zeta)  \sin\phi \sin^2\theta  \label{Eq_Ex_SF} \, ,
\end{eqnarray}
with $F_\alpha \equiv \alpha \omega m / \hbar$. We realize that the ISHE $(A)$ term plays only a minor role for the total electric field, $\langle E_x \rangle_t =\langle E_x^{(A)} + E_x^{(B)} + E_x^{(C)}  \rangle_t$.  Note that  for $\zeta \simeq 0$ the inverse spin filter contribution is of the same order of magnitude as the SGE term, whereas it vanishes for $\zeta = 1$.

\subsection{Orthogonal measurement} \label{Subsec_Narrowy}
In the case of an orthogonal measurement, see Fig.~\ref{Fig_Fields}, r.h.s., the contribution given in Eq.~(\ref{Eq_F_SFy}) vanishes since the spin current $\bj_y$ lacks a contribution parallel to the magnetization ($l_y \ll \lsf$). For the DC electric field along $y$-direction, we find
\begin{eqnarray}
\left\langle E_y^\mathrm{(A)} \right\rangle_t  &\sim & \betaDP  \theta^\mathrm{sH} \frac{F_\alpha}{e} \ll \theta^\mathrm{sH} \frac{F_\alpha}{e} \, , \label{Eq_Ey_SGE} \\
\left\langle E_y^\mathrm{(B)} \right\rangle_t  &=& -2 \theta^\mathrm{sH}  \frac{F_\alpha}{e} (1 + \zeta) \cos \phi  \sin^2\theta \label{Eq_Ey_SF}  \, , \\
\left\langle E_y^\mathrm{(C)} \right\rangle_t  &=& 0 \, ,
\end{eqnarray}
leaving only the SGE term to contribute to the total DC electric field.

\subsection{Discussion}
\begin{figure}[ht]
\includegraphics[width=0.9\columnwidth]{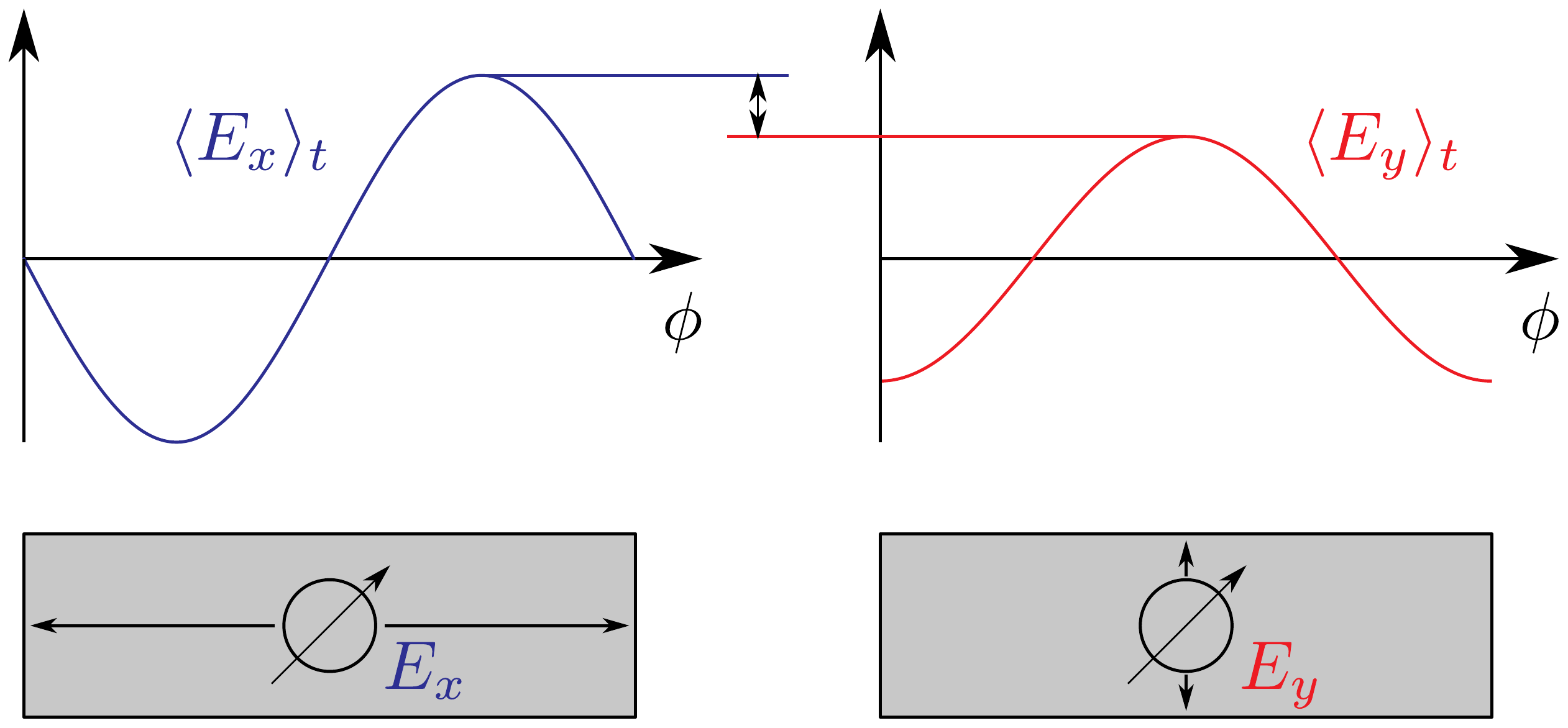}
\caption{Bottom part of figure: top view of the setup; the length is denoted $l_x$, the width $l_y$ ($l_y \ll l_x$). Top part of figure: qualitative plot of the DC electric fields, $\langle eE_x \rangle_t$ and $\langle eE_y \rangle_t$, for a longitudinal (left) and orthogonal (right) measurement. In both cases we set $\zeta = 0$. \label{Fig_Fields}}
\end{figure} 

Comparing the results for the longitudinal and the orthogonal measurement, we see that the signal can be up to $1.5$ times larger in the longitudinal measurement (for $\zeta=0$), see Fig.~\ref{Fig_Fields}. For a 2D electron gas one
should thus be able to probe a Rashba-induced SGE, 
and by comparing samples of different thicknesses, to additionally obtain estimates of $\alpha$ and $\zeta$.

Recent articles\cite{tserkovnyak2002,tserkovnyak2005,mosendz2010_2,obstbaum2014} discussed spin pumping and the induced ISHE on the basis of a spin current $\bj_z$ which flows perpendicular to the interface, i.e., in $z$-direction into the normal-metal film. This is significantly different from the situation we are discussing here ($\bj_z=0$). Nevertheless, the electric field estimated in such a way\cite{mosendz2010_2} shows the same angular-dependence of the magnetization as our result:
\begin{eqnarray}
\textnormal{Ref.~\onlinecite{mosendz2010_2}} & \Rightarrow & E_x \sim F_{g^{\uparrow\downarrow}} \sin \phi \sin^2 \theta \, , \\
\textnormal{Eq.~(\ref{Eq_Ex_SGE})}  & \Rightarrow & E_x \sim F_\alpha \sin \phi \sin^2 \theta \, .
\end{eqnarray}
Comparing the relevant forces, $F_{g^{\uparrow\downarrow}} = e^2\omega g^{\uparrow\downarrow}/(4\sigma_D)$ and $F_\alpha$, for reasonable parameter values, $g^{\uparrow\downarrow}\approx\SI{2.1e19}{m^{-2}}$ and $\sigma_D \approx\SI{2.4e6}{\ohm^{-1}.m^{-1}}$ for a Pt film,\cite{mosendz2010_2} we find the forces to be of the same order of magnitude for $\alpha \approx \SI{0.3}{\electronvolt.\angstrom}$. Thus we conclude that the SGE contribution and the inverse spin filter effects due to Rashba-induced spin currents and spin density, as discussed here, should both be taken into account when interpreting experiments.

%%%%%%%%%%%%%%%%%%%%%%%%%%%%%%%%%%%%%%%%%%%%%%%%%%%%%%%%%%%%%%%%%%%%%%%%%%%% CONCLUSION %%%%%%%%%%%%%%%%%%%%%%%%%%%%%%%%%%%%%%%%%%%%%%%%%%%%%%%%%%%%%%%%%%%%%%%%%

\section{Conclusions}
\label{Sec_conclusions}
We have studied the spin-charge coupled dynamics in a magnetized thin metallic film with Rashba spin-orbit coupling. 
In particular, we have considered a generalized Elliott-Yafet collision integral, valid for arbitrary spin-orbit fields
and magnetic textures, and taken into account anisotropic spin-flip processes. {Significant modifications of the kinetic 
equations describing spin and charge transport have been found. The effective force acting on the charge carriers in the presence of
spin-orbit coupling has been derived in a very general form, and evaluated in detail for the case of a time-dependent
texture.}
For a narrow wire in the typical spin pumping configuration an in-plane electric field is generated, for which
the spin galvanic effect is crucial, while the (in-plane) spin Hall effect turns out to be negligible.  
However, an additional contribution of similar strength from an `inverse spin filter' effect is found to be relevant 
for a longitudinal measurement, while it vanishes for an orthogonal measurement.
This suggests the possibility of determining the strength of the spin galvanic effect and the spin-orbit coupling parameter---Rashba in our 
specific scenario---by performing both measurements on the same sample.

\begin{acknowledgments}
We acknowledge stimulating discussions with C.\ Back, L.\ Chen, M.\ Decker, and R.\ Raimondi.  We especially are indebted to L.\ Chen for helpful 
comments on the manuscript.  
We acknowledge financial support from the German Research Foundation (DFG) through TRR 80 (UE, ST)
and SFB 689 (CG).
\end{acknowledgments}

%%%%%%%%%%%%%%%%%%%%%%%%%%%%%%%%%%%%%%%%%%%%%%%%%%%%%%%%%%%%%%%%%%%%%%%%%%%% APPENDIX %%%%%%%%%%%%%%%%%%%%%%%%%%%%%%%%%%%%%%%%%%%%%%%%%%%%%%%%%%%%%%%%%%%%%%%%%

\appendix

\section{The Elliott-Yafet collision operator}
\label{App_EY}
In this appendix we follow the procedure outlined in Ref.~\onlinecite{gorini2010}. We start by deriving the Elliott-Yafet collision operator within first order in the SU(2) fields from the microscopic Green's function $G$ and the Elliott-Yafet self-energy $\Sigma_\mathrm{EY}$ (diagrammatically depicted in Fig.~\ref{Fig_selfenergy}) in the two-dimensional case, and comment on the three-dimensional case at the end. 

The Elliott-Yafet collision operator is given by
\begin{equation} \label{EqAPP_colldef}
I_\mathrm{EY} = -\frac{1}{4\pi\hbar} \int d\epsilon  \tilde{L}^K \, ,
\end{equation}
with $L = \left[\Sigma_\mathrm{EY} , G\right]$. The superscript $K$ represents the Keldysh component, and the tilde the SU(2) shift; clearly $\tilde{L}=[{\tilde\Sigma}_\mathrm{EY} , \tilde{G}]$. In first order, we have
\begin{equation}
\tilde{L} \approx L - \frac{1}{2} \left\lbrace A_\mu , \partial_p^\mu L \right\rbrace \, , \label{EqAPP_shift}
\end{equation}
with the four-potential $A_\mu= (-\Psi, \bcalA)$ and $\partial_p^\mu = (\partial_\epsilon, \nabla_\bp)$. We recall that the components of the four-potential are SU(2) gauge fields, i.e., $\Psi = \Psi^a \sigma^a/2$ and $\bcalA = \bcalA^a \sigma^a/2$. In order to derive the explicit expression for the collision operator, we need in the first step the impurity averaged and SU(2) shifted self-energy, compare Fig.~\ref{Fig_selfenergy}, which reads
\begin{equation}
\tilde{\Sigma}_\mathrm{EY} = \tilde{\Sigma}_\mathrm{EY}^0 + \delta \tilde{\Sigma}_\mathrm{EY}^\Psi + \delta \tilde{\Sigma}_\mathrm{EY}^{\bcalA} \, ,
\end{equation}
with
\begin{widetext}
\begin{eqnarray}
 \tilde{\Sigma}_{\mathrm{EY}}^0 &=& C \int \frac{d^2 p'}{(2\pi\hbar)^2} \sigma^z \tilde{G}(\bp')\sigma^z \left( \bp \times \bp' \right)_z^2 \label{EqAPP_EX0} ,\\
 \delta \tilde{\Sigma}_{\mathrm{EY}}^\Psi &=& \frac{C}{2} \partial_\epsilon \hspace*{-2pt}\int \frac{d^2 p'}{(2\pi\hbar)^2}  \bigg(\sigma^z \left\lbrace \Psi^a \frac{\sigma^a}{2} , \tilde{G}(\bp')  \right\rbrace \sigma^z  - \left\lbrace \Psi^a \frac{\sigma^a}{2} ,\sigma^z \tilde{G}(\bp') \sigma^z \right\rbrace \bigg) \left( \bp \times \bp' \right)_z^2 ,\\
 \delta \tilde{\Sigma}_{\mathrm{EY}}^{\bcalA} &=& \frac{C}{2} \int \frac{d^2 p'}{(2\pi\hbar)^2} \bigg(\sigma^z \left\lbrace \bcalA^a \frac{\sigma^a}{2} , [\nabla_{\bp'}\tilde{G}(\bp')]  \right\rbrace \sigma^z  -  \left\lbrace \bcalA^a \frac{\sigma^a}{2} ,\sigma^z \tilde{G}(\bp') \sigma^z \right\rbrace \nabla_{\bp} \bigg) \left( \bp \times \bp' \right)_z^2  , \label{EqAPP_EXA}
\end{eqnarray}
\end{widetext}
where $C = (2\pi \tau \hbar^3 N_0)^{-1}(\lambda/2)^4$.
For the Green's functions we have
\begin{equation}
 \tilde{G}^K = \left( \tilde{G}^R - \tilde{G}^A \right)(1-2f) \, ,
\end{equation}
where $f=f^0 + {\bf f} \cdot \bsigma$ denotes the distribution function $2\times2$ matrix; furthermore,
\begin{equation}
\tilde{G}^R - \tilde{G}^A = -2\pi i \delta (\epsilon - \epsilon_\bp) \, . \label{EqAPP_GRGA}
\end{equation}
Note that the SU(2) shifted retarded and advanced Green's functions are diagonal in spin. Inserting $\tilde{L}$ into Eq.~(\ref{EqAPP_colldef}) and using Eqs.~(\ref{EqAPP_EX0})--(\ref{EqAPP_GRGA}) leads to the collision operators as given in Eqs.~(\ref{Eq_EY})--(\ref{Eq_EY2}) for $\zeta = 0$, corresponding to the 2D case. 

When we consider the bulk 3D case we have the following replacement:
\begin{equation}
\sigma^z\dots\sigma^z \left( \bp \times \bp' \right)_z^2
\rightarrow
\sigma^a\dots\sigma^b\left( \bp \times \bp' \right)_a\left( \bp \times \bp' \right)_b
\end{equation}
within the integrals in Eqs.~(\ref{EqAPP_EX0})--(\ref{EqAPP_EXA}), respectively. Then we obtain, by the same procedure as outlined in Sec.~\ref{Sec_intro}, Eqs.~(\ref{Eq_EY})--(\ref{Eq_EY2}), for $\zeta=1$, except for a small numerical difference related to the angular average in 2D versus 3D.
In order to describe the anisotropy in the intermediate regime, we insert the matrix $\Gamma = \mathrm{diag}(1,1,\zeta)$ with $0\leq \zeta \leq 1$; 
cf.\ Eqs.\ (\ref{Eq_EY}) and (\ref{Eq_EY1}).

\section{Narrow wires}
\label{App_Narrow}
Here we show that the contribution which is polarized parallel to the magnetization of the spin current flowing in the narrow direction vanishes in the homogeneous case. We consider a narrow wire along $x$-direction, i.e., $l_y \ll \lsf$ and $l_x \gg \lsf$, and an open circuit condition, $\bj_y(y=0)=\bj_y(y=l_y)=0$. For the sake of simplicity we put $\zeta=1$.  Since the system is homogeneous, we assume that the spin current flowing in $x$-direction is homogeneous as well, and given by Eqs.~(\ref{Eq_spincurrent})--(\ref{Eq_sc_deltas}) with $\tilde{\nabla}_x \rightarrow [\ba_x]_\times$. According to Eqs.~(\ref{Eq_ssplit})--(\ref{Eq_ssc}) the spin density can be expressed as
\begin{equation} \label{Eq_spindensity_ydep}
\delta \bs(y) = \delta \bs_0 - \tausf M_s^{-1} \tilde{\nabla}_y \bj_y(y) 
\end{equation}
with 
\begin{equation}
\delta \bs_0 = \left( \delta \bs \right)_\bn - \tausf M_s^{-1} [\ba_x]_\times \bj_x \, .
\end{equation}
In addition, $M_s = M(\tau\rightarrow\tausf)$, where $M$, cf.\ Eq.~(\ref{Eq_M}), is explicitly given by
\begin{equation}
M=1+ \frac{\Exc\tau}{\hbar}[\bn]_\times =
\betatau^{-1}
\begin{pmatrix}
\betatau & -n_z & n_y \\
n_z & \betatau & -n_x \\
-n_y & n_x & \betatau
\end{pmatrix} \, ,
\end{equation}
where $\betatau=\hbar/\Exc \tau$.
The spin current flowing in $y$-direction is given by Eqs.~(\ref{Eq_spincurrent})--(\ref{Eq_sc_deltas}):
\begin{equation} \label{Eq_spincurrent_ydep}
\bj_y(y) = D M^{-1} \tilde{\nabla}_y M^{-1} \left( N_0 \betatau^{-1} \dot{\bn} - \delta \bs(y) \right) \, .
\end{equation}
We insert Eq.~(\ref{Eq_spindensity_ydep}) into Eq.~(\ref{Eq_spincurrent_ydep}) and obtain approximately ($\tilde{\nabla}_y \bj_y \simeq \nabla_y \bj_y$) the following differential equation:
\begin{equation} \label{Eq_diffEq}
\left( 1 -\lsf^2 M_s^{-1} M^{-2}  \nabla_y^2  \right) \bj_y(y) =  \bj_{y,0} \, ,
\end{equation}
where the r.h.s., which is spatially constant, is given by
\begin{equation}
\bj_{y,0} = D M^{-1} [\ba_y]_\times M^{-1} \left( N_0 \betatau^{-1} \dot{\bn} - \delta \bs_0 \right) \, .
\end{equation}
It is apparent that $\bj_{y,0}$  is a particular solution of Eq.~(\ref{Eq_diffEq}). In order to determine the complete solution $\bj_y = \bj_{y,h} + \bj_{y,0}$, we have to add the solution of the homogeneous differential equation, which can be written as follows:
\begin{equation} \label{Eq_diffEqhom}
\left( M_s M^2 -  \lsf^2 \nabla_y^2  \right) \bj_{y,h}(y) = 0 \, .
\end{equation}
It is convenient to change the basis by the following transformation:
\begin{equation}
\mathbb{R} = \begin{pmatrix}
n_x & \dot{n}_x / |\dot{\bn}| & (\bn\times\dot{\bn})_x / |\dot{\bn}| \\
n_y & \dot{n}_y / |\dot{\bn}| & (\bn\times\dot{\bn})_y / |\dot{\bn}| \\
n_z & \dot{n}_z / |\dot{\bn}| & (\bn\times\dot{\bn})_z / |\dot{\bn}|
\end{pmatrix} \, ,
\end{equation}
which replaces $\bn$ by ${\bf e}_x$ in Eq.~(\ref{Eq_diffEqhom}):
\begin{equation} \label{Eq_diffEqhomtrans}
\left[ \left(1 +  \betasf^{-1} [{\bf e}_x]_\times \right) \left(1 + \betatau^{-1} [{\bf e}_x]_\times\right)^2 - \lsf^2 \nabla_y^2  \right]  \boldsymbol{\tilde{\jmath}}_{y,h}(y) = 0 ,
\end{equation}
with $\boldsymbol{\tilde{\jmath}}_{y,h} = \mathbb{R}^T \bj_{y,h}$. The $x$-component of $\boldsymbol{\tilde{\jmath}}_{y,h}$ represents the contribution of the spin current which is parallel to the magnetization. The matrix in Eq.~(\ref{Eq_diffEqhomtrans}) has the eigenvalue $1$ with eigenvector ${\bf e}_x$, leading to
\begin{equation}
\tilde{\jmath}_{y,h}^x = A \exp\left(-\frac{y}{\lsf}\right) + B \exp\left(+\frac{y}{\lsf}\right) \, .
\end{equation}
For $l_y \ll \lsf$ the general solution thus reads
\begin{equation} \label{Eq_jyx}
\tilde{\jmath}_y^x  = A \left( 1- \frac{y}{\lsf} \right) + B \left( 1+ \frac{y}{\lsf} \right) + \tilde{\jmath}_{y,0}^x \, ,
\end{equation}
where $ \tilde{\jmath}_{y,0}^x$ is the $x$-component of $\mathbb{R}^T \bj_{y,0}$.
We then use the boundary conditions, $\bj_y(y=0)=\bj_y(y=l_y)=0$, with the result
\begin{equation}
A = B = - \frac{1}{2} \tilde{\jmath}_{y,0}^x \, .
\end{equation}
Finally we insert $A$ and $B$ into Eq.~(\ref{Eq_jyx}) and find $\tilde{\jmath}_y^x = 0$, therefore the spin current contribution which is parallel to the magnetization vanishes, $\bn \cdot \bj_y = 0$. 

We remark that the transverse-polarization components of the spin current, i.e. $\tilde{\jmath}_y^y$ and $\tilde{\jmath}_y^z$, do not vanish since the transverse spin relaxation length $\lsf^\perp$ is orders of magnitude smaller than $\lsf$.\cite{asaya2012}  
An explicit solution of the above diffusion equations, obtained assuming $\Exc\tau/\hbar\gg1$, 
yields $\lsf^\perp < \lsf \betatau$.

\bibliography{paper_magnet_tex_biblio}

\begin{thebibliography}{63}%
\makeatletter
\providecommand \@ifxundefined [1]{%
 \@ifx{#1\undefined}
}%
\providecommand \@ifnum [1]{%
 \ifnum #1\expandafter \@firstoftwo
 \else \expandafter \@secondoftwo
 \fi
}%
\providecommand \@ifx [1]{%
 \ifx #1\expandafter \@firstoftwo
 \else \expandafter \@secondoftwo
 \fi
}%
\providecommand \natexlab [1]{#1}%
\providecommand \enquote  [1]{``#1''}%
\providecommand \bibnamefont  [1]{#1}%
\providecommand \bibfnamefont [1]{#1}%
\providecommand \citenamefont [1]{#1}%
\providecommand \href@noop [0]{\@secondoftwo}%
\providecommand \href [0]{\begingroup \@sanitize@url \@href}%
\providecommand \@href[1]{\@@startlink{#1}\@@href}%
\providecommand \@@href[1]{\endgroup#1\@@endlink}%
\providecommand \@sanitize@url [0]{\catcode `\\12\catcode `\$12\catcode
  `\&12\catcode `\#12\catcode `\^12\catcode `\_12\catcode `\%12\relax}%
\providecommand \@@startlink[1]{}%
\providecommand \@@endlink[0]{}%
\providecommand \url  [0]{\begingroup\@sanitize@url \@url }%
\providecommand \@url [1]{\endgroup\@href {#1}{\urlprefix }}%
\providecommand \urlprefix  [0]{URL }%
\providecommand \Eprint [0]{\href }%
\providecommand \doibase [0]{http://dx.doi.org/}%
\providecommand \selectlanguage [0]{\@gobble}%
\providecommand \bibinfo  [0]{\@secondoftwo}%
\providecommand \bibfield  [0]{\@secondoftwo}%
\providecommand \translation [1]{[#1]}%
\providecommand \BibitemOpen [0]{}%
\providecommand \bibitemStop [0]{}%
\providecommand \bibitemNoStop [0]{.\EOS\space}%
\providecommand \EOS [0]{\spacefactor3000\relax}%
\providecommand \BibitemShut  [1]{\csname bibitem#1\endcsname}%
\let\auto@bib@innerbib\@empty
%</preamble>
\bibitem [{\citenamefont {$\check{\textnormal{Z}}$uti\'c}\ \emph
  {et~al.}(2004)\citenamefont {$\check{\textnormal{Z}}$uti\'c}, \citenamefont
  {Fabian},\ and\ \citenamefont {Das~Sarma}}]{zutic2004}%
  \BibitemOpen
  \bibfield  {author} {\bibinfo {author} {\bibfnamefont {I.}~\bibnamefont
  {$\check{\textnormal{Z}}$uti\'c}}, \bibinfo {author} {\bibfnamefont
  {J.}~\bibnamefont {Fabian}}, \ and\ \bibinfo {author} {\bibfnamefont
  {S.}~\bibnamefont {Das~Sarma}},\ }\href {\doibase 10.1103/RevModPhys.76.323}
  {\bibfield  {journal} {\bibinfo  {journal} {Rev. Mod. Phys.}\ }\textbf
  {\bibinfo {volume} {76}},\ \bibinfo {pages} {323} (\bibinfo {year}
  {2004})}\BibitemShut {NoStop}%
\bibitem [{\citenamefont {Urban}\ \emph {et~al.}(2001)\citenamefont {Urban},
  \citenamefont {Woltersdorf},\ and\ \citenamefont {Heinrich}}]{urban2001}%
  \BibitemOpen
  \bibfield  {author} {\bibinfo {author} {\bibfnamefont {R.}~\bibnamefont
  {Urban}}, \bibinfo {author} {\bibfnamefont {G.}~\bibnamefont {Woltersdorf}},
  \ and\ \bibinfo {author} {\bibfnamefont {B.}~\bibnamefont {Heinrich}},\
  }\href {\doibase 10.1103/PhysRevLett.87.217204} {\bibfield  {journal}
  {\bibinfo  {journal} {Phys. Rev. Lett.}\ }\textbf {\bibinfo {volume} {87}},\
  \bibinfo {pages} {217204} (\bibinfo {year} {2001})}\BibitemShut {NoStop}%
\bibitem [{\citenamefont {Tserkovnyak}\ \emph {et~al.}(2002)\citenamefont
  {Tserkovnyak}, \citenamefont {Brataas},\ and\ \citenamefont
  {Bauer}}]{tserkovnyak2002}%
  \BibitemOpen
  \bibfield  {author} {\bibinfo {author} {\bibfnamefont {Y.}~\bibnamefont
  {Tserkovnyak}}, \bibinfo {author} {\bibfnamefont {A.}~\bibnamefont
  {Brataas}}, \ and\ \bibinfo {author} {\bibfnamefont {G.~E.~W.}\ \bibnamefont
  {Bauer}},\ }\href {\doibase 10.1103/PhysRevLett.88.117601} {\bibfield
  {journal} {\bibinfo  {journal} {Phys. Rev. Lett.}\ }\textbf {\bibinfo
  {volume} {88}},\ \bibinfo {pages} {117601} (\bibinfo {year}
  {2002})}\BibitemShut {NoStop}%
\bibitem [{\citenamefont {Azevedo}\ \emph {et~al.}(2011)\citenamefont
  {Azevedo}, \citenamefont {Vilela-Le\~ao}, \citenamefont
  {Rodr\'{\i}guez-Su\'arez}, \citenamefont {Lacerda~Santos},\ and\
  \citenamefont {Rezende}}]{azevedo2011}%
  \BibitemOpen
  \bibfield  {author} {\bibinfo {author} {\bibfnamefont {A.}~\bibnamefont
  {Azevedo}}, \bibinfo {author} {\bibfnamefont {L.~H.}\ \bibnamefont
  {Vilela-Le\~ao}}, \bibinfo {author} {\bibfnamefont {R.~L.}\ \bibnamefont
  {Rodr\'{\i}guez-Su\'arez}}, \bibinfo {author} {\bibfnamefont {A.~F.}\
  \bibnamefont {Lacerda~Santos}}, \ and\ \bibinfo {author} {\bibfnamefont
  {S.~M.}\ \bibnamefont {Rezende}},\ }\href {\doibase
  https://doi.org/10.1103/PhysRevB.83.144402} {\bibfield  {journal} {\bibinfo
  {journal} {Phys. Rev. B}\ }\textbf {\bibinfo {volume} {83}},\ \bibinfo
  {pages} {144402} (\bibinfo {year} {2011})}\BibitemShut {NoStop}%
\bibitem [{\citenamefont {Czeschka}\ \emph {et~al.}(2011)\citenamefont
  {Czeschka}, \citenamefont {Dreher}, \citenamefont {Brandt}, \citenamefont
  {Weiler}, \citenamefont {Althammer}, \citenamefont {Imort}, \citenamefont
  {Reiss}, \citenamefont {Thomas}, \citenamefont {Schoch}, \citenamefont
  {Limmer}, \citenamefont {Huebl}, \citenamefont {Gross},\ and\ \citenamefont
  {Goennenwein}}]{czeschka2011}%
  \BibitemOpen
  \bibfield  {author} {\bibinfo {author} {\bibfnamefont {F.~D.}\ \bibnamefont
  {Czeschka}}, \bibinfo {author} {\bibfnamefont {L.}~\bibnamefont {Dreher}},
  \bibinfo {author} {\bibfnamefont {M.~S.}\ \bibnamefont {Brandt}}, \bibinfo
  {author} {\bibfnamefont {M.}~\bibnamefont {Weiler}}, \bibinfo {author}
  {\bibfnamefont {M.}~\bibnamefont {Althammer}}, \bibinfo {author}
  {\bibfnamefont {I.-M.}\ \bibnamefont {Imort}}, \bibinfo {author}
  {\bibfnamefont {G.}~\bibnamefont {Reiss}}, \bibinfo {author} {\bibfnamefont
  {A.}~\bibnamefont {Thomas}}, \bibinfo {author} {\bibfnamefont
  {W.}~\bibnamefont {Schoch}}, \bibinfo {author} {\bibfnamefont
  {W.}~\bibnamefont {Limmer}}, \bibinfo {author} {\bibfnamefont
  {H.}~\bibnamefont {Huebl}}, \bibinfo {author} {\bibfnamefont
  {R.}~\bibnamefont {Gross}}, \ and\ \bibinfo {author} {\bibfnamefont
  {S.~T.~B.}\ \bibnamefont {Goennenwein}},\ }\href {\doibase
  10.1103/PhysRevLett.107.046601} {\bibfield  {journal} {\bibinfo  {journal}
  {Phys. Rev. Lett.}\ }\textbf {\bibinfo {volume} {107}},\ \bibinfo {pages}
  {046601} (\bibinfo {year} {2011})}\BibitemShut {NoStop}%
\bibitem [{\citenamefont {Saitoh}\ \emph {et~al.}(2006)\citenamefont {Saitoh},
  \citenamefont {Ueda}, \citenamefont {Miyajima},\ and\ \citenamefont
  {Tatara}}]{saitoh2006}%
  \BibitemOpen
  \bibfield  {author} {\bibinfo {author} {\bibfnamefont {E.}~\bibnamefont
  {Saitoh}}, \bibinfo {author} {\bibfnamefont {M.}~\bibnamefont {Ueda}},
  \bibinfo {author} {\bibfnamefont {H.}~\bibnamefont {Miyajima}}, \ and\
  \bibinfo {author} {\bibfnamefont {G.}~\bibnamefont {Tatara}},\ }\href
  {\doibase http://dx.doi.org/10.1063/1.2199473} {\bibfield  {journal}
  {\bibinfo  {journal} {Appl. Phys. Lett.}\ }\textbf {\bibinfo {volume} {88}},\
  \bibinfo {pages} {182509} (\bibinfo {year} {2006})}\BibitemShut {NoStop}%
\bibitem [{\citenamefont {Kimura}\ \emph {et~al.}(2007)\citenamefont {Kimura},
  \citenamefont {Otani}, \citenamefont {Sato}, \citenamefont {Takahashi},\ and\
  \citenamefont {Maekawa}}]{kimura2007}%
  \BibitemOpen
  \bibfield  {author} {\bibinfo {author} {\bibfnamefont {T.}~\bibnamefont
  {Kimura}}, \bibinfo {author} {\bibfnamefont {Y.}~\bibnamefont {Otani}},
  \bibinfo {author} {\bibfnamefont {T.}~\bibnamefont {Sato}}, \bibinfo {author}
  {\bibfnamefont {S.}~\bibnamefont {Takahashi}}, \ and\ \bibinfo {author}
  {\bibfnamefont {S.}~\bibnamefont {Maekawa}},\ }\href {\doibase
  10.1103/PhysRevLett.98.156601} {\bibfield  {journal} {\bibinfo  {journal}
  {Phys. Rev. Lett.}\ }\textbf {\bibinfo {volume} {98}},\ \bibinfo {pages}
  {156601} (\bibinfo {year} {2007})}\BibitemShut {NoStop}%
\bibitem [{\citenamefont {Mosendz}\ \emph {et~al.}(2010)\citenamefont
  {Mosendz}, \citenamefont {Vlaminck}, \citenamefont {Pearson}, \citenamefont
  {Fradin}, \citenamefont {Bauer}, \citenamefont {Bader},\ and\ \citenamefont
  {Hoffmann}}]{mosendz2010_2}%
  \BibitemOpen
  \bibfield  {author} {\bibinfo {author} {\bibfnamefont {O.}~\bibnamefont
  {Mosendz}}, \bibinfo {author} {\bibfnamefont {V.}~\bibnamefont {Vlaminck}},
  \bibinfo {author} {\bibfnamefont {J.~E.}\ \bibnamefont {Pearson}}, \bibinfo
  {author} {\bibfnamefont {F.~Y.}\ \bibnamefont {Fradin}}, \bibinfo {author}
  {\bibfnamefont {G.~E.~W.}\ \bibnamefont {Bauer}}, \bibinfo {author}
  {\bibfnamefont {S.~D.}\ \bibnamefont {Bader}}, \ and\ \bibinfo {author}
  {\bibfnamefont {A.}~\bibnamefont {Hoffmann}},\ }\href {\doibase
  10.1103/PhysRevB.82.214403} {\bibfield  {journal} {\bibinfo  {journal} {Phys.
  Rev. B}\ }\textbf {\bibinfo {volume} {82}},\ \bibinfo {pages} {214403}
  (\bibinfo {year} {2010})}\BibitemShut {NoStop}%
\bibitem [{\citenamefont {Obstbaum}\ \emph {et~al.}(2014)\citenamefont
  {Obstbaum}, \citenamefont {H\"{a}rtinger}, \citenamefont {Bauer},
  \citenamefont {Meier}, \citenamefont {Swientek}, \citenamefont {Back},\ and\
  \citenamefont {Woltersdorf}}]{obstbaum2014}%
  \BibitemOpen
  \bibfield  {author} {\bibinfo {author} {\bibfnamefont {M.}~\bibnamefont
  {Obstbaum}}, \bibinfo {author} {\bibfnamefont {M.}~\bibnamefont
  {H\"{a}rtinger}}, \bibinfo {author} {\bibfnamefont {H.~G.}\ \bibnamefont
  {Bauer}}, \bibinfo {author} {\bibfnamefont {T.}~\bibnamefont {Meier}},
  \bibinfo {author} {\bibfnamefont {F.}~\bibnamefont {Swientek}}, \bibinfo
  {author} {\bibfnamefont {C.~H.}\ \bibnamefont {Back}}, \ and\ \bibinfo
  {author} {\bibfnamefont {G.}~\bibnamefont {Woltersdorf}},\ }\href {\doibase
  http://dx.doi.org/10.1103/PhysRevB.89.060407} {\bibfield  {journal} {\bibinfo
   {journal} {Phys. Rev. B}\ }\textbf {\bibinfo {volume} {89}},\ \bibinfo
  {pages} {060407(R)} (\bibinfo {year} {2014})}\BibitemShut {NoStop}%
\bibitem [{Note1()}]{Note1}%
  \BibitemOpen
  \bibinfo {note} {Magnetic insulators, ferro- or ferrimagnets are used as
  `spin pumpers' in different experiments. For our purposes the difference
  between them is actually irrelevant, and we will thus speak generally about
  `magnets' or `magnetic materials'.}\BibitemShut {Stop}%
\bibitem [{\citenamefont {Seki}\ \emph {et~al.}(2008)\citenamefont {Seki},
  \citenamefont {Hasegawa}, \citenamefont {Mitani}, \citenamefont {Takahashi},
  \citenamefont {Imamura}, \citenamefont {Maekawa}, \citenamefont {Nitta},\
  and\ \citenamefont {Takanashi}}]{seki2008}%
  \BibitemOpen
  \bibfield  {author} {\bibinfo {author} {\bibfnamefont {T.}~\bibnamefont
  {Seki}}, \bibinfo {author} {\bibfnamefont {Y.}~\bibnamefont {Hasegawa}},
  \bibinfo {author} {\bibfnamefont {S.}~\bibnamefont {Mitani}}, \bibinfo
  {author} {\bibfnamefont {S.}~\bibnamefont {Takahashi}}, \bibinfo {author}
  {\bibfnamefont {H.}~\bibnamefont {Imamura}}, \bibinfo {author} {\bibfnamefont
  {S.}~\bibnamefont {Maekawa}}, \bibinfo {author} {\bibfnamefont
  {J.}~\bibnamefont {Nitta}}, \ and\ \bibinfo {author} {\bibfnamefont
  {K.}~\bibnamefont {Takanashi}},\ }\href {\doibase 10.1038/nmat2098}
  {\bibfield  {journal} {\bibinfo  {journal} {Nat. Mater.}\ }\textbf {\bibinfo
  {volume} {7}},\ \bibinfo {pages} {125} (\bibinfo {year} {2008})}\BibitemShut
  {NoStop}%
\bibitem [{\citenamefont {Vila}\ \emph {et~al.}(2007)\citenamefont {Vila},
  \citenamefont {Kimura},\ and\ \citenamefont {Otani}}]{vila2007}%
  \BibitemOpen
  \bibfield  {author} {\bibinfo {author} {\bibfnamefont {L.}~\bibnamefont
  {Vila}}, \bibinfo {author} {\bibfnamefont {T.}~\bibnamefont {Kimura}}, \ and\
  \bibinfo {author} {\bibfnamefont {Y.~C.}\ \bibnamefont {Otani}},\ }\href
  {\doibase 10.1103/PhysRevLett.99.226604} {\bibfield  {journal} {\bibinfo
  {journal} {Phys. Rev. Lett.}\ }\textbf {\bibinfo {volume} {99}},\ \bibinfo
  {pages} {226604} (\bibinfo {year} {2007})}\BibitemShut {NoStop}%
\bibitem [{\citenamefont {Liu}\ \emph {et~al.}(2011)\citenamefont {Liu},
  \citenamefont {Moriyama}, \citenamefont {Ralph},\ and\ \citenamefont
  {Buhrman}}]{liu2011}%
  \BibitemOpen
  \bibfield  {author} {\bibinfo {author} {\bibfnamefont {L.}~\bibnamefont
  {Liu}}, \bibinfo {author} {\bibfnamefont {T.}~\bibnamefont {Moriyama}},
  \bibinfo {author} {\bibfnamefont {D.~C.}\ \bibnamefont {Ralph}}, \ and\
  \bibinfo {author} {\bibfnamefont {R.~A.}\ \bibnamefont {Buhrman}},\ }\href
  {\doibase 10.1103/PhysRevLett.106.036601} {\bibfield  {journal} {\bibinfo
  {journal} {Phys. Rev. Lett.}\ }\textbf {\bibinfo {volume} {106}},\ \bibinfo
  {pages} {036601} (\bibinfo {year} {2011})}\BibitemShut {NoStop}%
\bibitem [{\citenamefont {Hahn}\ \emph {et~al.}(2013)\citenamefont {Hahn},
  \citenamefont {de~Loubens}, \citenamefont {Klein}, \citenamefont {Viret},
  \citenamefont {Naletov},\ and\ \citenamefont {Ben~Youssef}}]{hahn2013}%
  \BibitemOpen
  \bibfield  {author} {\bibinfo {author} {\bibfnamefont {C.}~\bibnamefont
  {Hahn}}, \bibinfo {author} {\bibfnamefont {G.}~\bibnamefont {de~Loubens}},
  \bibinfo {author} {\bibfnamefont {O.}~\bibnamefont {Klein}}, \bibinfo
  {author} {\bibfnamefont {M.}~\bibnamefont {Viret}}, \bibinfo {author}
  {\bibfnamefont {V.~V.}\ \bibnamefont {Naletov}}, \ and\ \bibinfo {author}
  {\bibfnamefont {J.}~\bibnamefont {Ben~Youssef}},\ }\href {\doibase
  10.1103/PhysRevB.87.174417} {\bibfield  {journal} {\bibinfo  {journal} {Phys.
  Rev. B}\ }\textbf {\bibinfo {volume} {87}},\ \bibinfo {pages} {174417}
  (\bibinfo {year} {2013})}\BibitemShut {NoStop}%
\bibitem [{\citenamefont {Liu}\ \emph {et~al.}(2012)\citenamefont {Liu},
  \citenamefont {Pai}, \citenamefont {Li}, \citenamefont {Tseng}, \citenamefont
  {Ralph},\ and\ \citenamefont {Buhrman}}]{liu2012}%
  \BibitemOpen
  \bibfield  {author} {\bibinfo {author} {\bibfnamefont {L.}~\bibnamefont
  {Liu}}, \bibinfo {author} {\bibfnamefont {C.-F.}\ \bibnamefont {Pai}},
  \bibinfo {author} {\bibfnamefont {Y.}~\bibnamefont {Li}}, \bibinfo {author}
  {\bibfnamefont {H.~W.}\ \bibnamefont {Tseng}}, \bibinfo {author}
  {\bibfnamefont {D.~C.}\ \bibnamefont {Ralph}}, \ and\ \bibinfo {author}
  {\bibfnamefont {R.~A.}\ \bibnamefont {Buhrman}},\ }\href {\doibase
  10.1126/science.1218197} {\bibfield  {journal} {\bibinfo  {journal}
  {Science}\ }\textbf {\bibinfo {volume} {336}},\ \bibinfo {pages} {555}
  (\bibinfo {year} {2012})}\BibitemShut {NoStop}%
\bibitem [{\citenamefont {Ando}\ \emph {et~al.}(2008)\citenamefont {Ando},
  \citenamefont {Takahashi}, \citenamefont {Harii}, \citenamefont {Sasage},
  \citenamefont {Ieda}, \citenamefont {Maekawa},\ and\ \citenamefont
  {Saitoh}}]{ando2008}%
  \BibitemOpen
  \bibfield  {author} {\bibinfo {author} {\bibfnamefont {K.}~\bibnamefont
  {Ando}}, \bibinfo {author} {\bibfnamefont {S.}~\bibnamefont {Takahashi}},
  \bibinfo {author} {\bibfnamefont {K.}~\bibnamefont {Harii}}, \bibinfo
  {author} {\bibfnamefont {K.}~\bibnamefont {Sasage}}, \bibinfo {author}
  {\bibfnamefont {J.}~\bibnamefont {Ieda}}, \bibinfo {author} {\bibfnamefont
  {S.}~\bibnamefont {Maekawa}}, \ and\ \bibinfo {author} {\bibfnamefont
  {E.}~\bibnamefont {Saitoh}},\ }\href {\doibase
  http://dx.doi.org/10.1103/PhysRevLett.101.036601} {\bibfield  {journal}
  {\bibinfo  {journal} {Phys. Rev. Lett.}\ }\textbf {\bibinfo {volume} {101}},\
  \bibinfo {pages} {036601} (\bibinfo {year} {2008})}\BibitemShut {NoStop}%
\bibitem [{\citenamefont {Pesin}\ and\ \citenamefont
  {MacDonald}(2012)}]{pesin2012}%
  \BibitemOpen
  \bibfield  {author} {\bibinfo {author} {\bibfnamefont {D.~A.}\ \bibnamefont
  {Pesin}}\ and\ \bibinfo {author} {\bibfnamefont {A.~H.}\ \bibnamefont
  {MacDonald}},\ }\href@noop {} {\bibfield  {journal} {\bibinfo  {journal}
  {Phys. Rev. B}\ }\textbf {\bibinfo {volume} {86}},\ \bibinfo {pages} {014416}
  (\bibinfo {year} {2012})}\BibitemShut {NoStop}%
\bibitem [{\citenamefont {Ganichev}\ \emph {et~al.}(2002)\citenamefont
  {Ganichev}, \citenamefont {Ivchenko}, \citenamefont {Bel'kov}, \citenamefont
  {Tarasenko}, \citenamefont {Sollinger}, \citenamefont {Weiss}, \citenamefont
  {Wegscheider},\ and\ \citenamefont {Prettl}}]{ganichev2002}%
  \BibitemOpen
  \bibfield  {author} {\bibinfo {author} {\bibfnamefont {S.~D.}\ \bibnamefont
  {Ganichev}}, \bibinfo {author} {\bibfnamefont {E.~L.}\ \bibnamefont
  {Ivchenko}}, \bibinfo {author} {\bibfnamefont {V.~V.}\ \bibnamefont
  {Bel'kov}}, \bibinfo {author} {\bibfnamefont {S.~A.}\ \bibnamefont
  {Tarasenko}}, \bibinfo {author} {\bibfnamefont {M.}~\bibnamefont
  {Sollinger}}, \bibinfo {author} {\bibfnamefont {D.}~\bibnamefont {Weiss}},
  \bibinfo {author} {\bibfnamefont {W.}~\bibnamefont {Wegscheider}}, \ and\
  \bibinfo {author} {\bibfnamefont {W.}~\bibnamefont {Prettl}},\ }\href
  {\doibase 10.1038/417153a} {\bibfield  {journal} {\bibinfo  {journal}
  {Nature}\ }\textbf {\bibinfo {volume} {417}},\ \bibinfo {pages} {153}
  (\bibinfo {year} {2002})}\BibitemShut {NoStop}%
\bibitem [{\citenamefont {Ganichev}\ \emph {et~al.}(2012)\citenamefont
  {Ganichev}, \citenamefont {Trushin},\ and\ \citenamefont
  {Schliemann}}]{ganichevreview2011}%
  \BibitemOpen
  \bibfield  {author} {\bibinfo {author} {\bibfnamefont {S.~D.}\ \bibnamefont
  {Ganichev}}, \bibinfo {author} {\bibfnamefont {M.}~\bibnamefont {Trushin}}, \
  and\ \bibinfo {author} {\bibfnamefont {J.}~\bibnamefont {Schliemann}},\ }in\
  \href@noop {} {\emph {\bibinfo {booktitle} {Handbook of Spin Transport and
  Magnetism}}},\ \bibinfo {editor} {edited by\ \bibinfo {editor} {\bibfnamefont
  {E.~Y.}\ \bibnamefont {Tsymbal}}\ and\ \bibinfo {editor} {\bibfnamefont
  {I.}~\bibnamefont {$\check{\textnormal{Z}}$uti\'c}}}\ (\bibinfo  {publisher}
  {CRC Press},\ \bibinfo {address} {Boca Raton, FL},\ \bibinfo {year} {2012})\
  pp.\ \bibinfo {pages} {487--495}\BibitemShut {NoStop}%
\bibitem [{\citenamefont {Raimondi}\ \emph {et~al.}(2006)\citenamefont
  {Raimondi}, \citenamefont {Gorini}, \citenamefont {Schwab},\ and\
  \citenamefont {Dzierzawa}}]{raimondi2006}%
  \BibitemOpen
  \bibfield  {author} {\bibinfo {author} {\bibfnamefont {R.}~\bibnamefont
  {Raimondi}}, \bibinfo {author} {\bibfnamefont {C.}~\bibnamefont {Gorini}},
  \bibinfo {author} {\bibfnamefont {P.}~\bibnamefont {Schwab}}, \ and\ \bibinfo
  {author} {\bibfnamefont {M.}~\bibnamefont {Dzierzawa}},\ }\href {\doibase
  10.1103/PhysRevB.74.035340} {\bibfield  {journal} {\bibinfo  {journal} {Phys.
  Rev. B}\ }\textbf {\bibinfo {volume} {74}},\ \bibinfo {pages} {035340}
  (\bibinfo {year} {2006})}\BibitemShut {NoStop}%
\bibitem [{\citenamefont {Borge}\ \emph {et~al.}(2014)\citenamefont {Borge},
  \citenamefont {Gorini}, \citenamefont {Vignale},\ and\ \citenamefont
  {Raimondi}}]{borge2014}%
  \BibitemOpen
  \bibfield  {author} {\bibinfo {author} {\bibfnamefont {J.}~\bibnamefont
  {Borge}}, \bibinfo {author} {\bibfnamefont {C.}~\bibnamefont {Gorini}},
  \bibinfo {author} {\bibfnamefont {G.}~\bibnamefont {Vignale}}, \ and\
  \bibinfo {author} {\bibfnamefont {R.}~\bibnamefont {Raimondi}},\ }\href
  {\doibase 10.1103/PhysRevB.89.245443} {\bibfield  {journal} {\bibinfo
  {journal} {Phys. Rev. B}\ }\textbf {\bibinfo {volume} {89}},\ \bibinfo
  {pages} {245443} (\bibinfo {year} {2014})}\BibitemShut {NoStop}%
\bibitem [{\citenamefont {Rojas~S\'anchez}\ \emph {et~al.}(2013)\citenamefont
  {Rojas~S\'anchez}, \citenamefont {Vila}, \citenamefont {Desfonds},
  \citenamefont {Gambarelli}, \citenamefont {Attan\'e}, \citenamefont
  {De~Teresa}, \citenamefont {Mag\'en},\ and\ \citenamefont
  {Fert}}]{sanchez2013}%
  \BibitemOpen
  \bibfield  {author} {\bibinfo {author} {\bibfnamefont {J.~C.}\ \bibnamefont
  {Rojas~S\'anchez}}, \bibinfo {author} {\bibfnamefont {L.}~\bibnamefont
  {Vila}}, \bibinfo {author} {\bibfnamefont {G.}~\bibnamefont {Desfonds}},
  \bibinfo {author} {\bibfnamefont {S.}~\bibnamefont {Gambarelli}}, \bibinfo
  {author} {\bibfnamefont {J.~P.}\ \bibnamefont {Attan\'e}}, \bibinfo {author}
  {\bibfnamefont {J.~M.}\ \bibnamefont {De~Teresa}}, \bibinfo {author}
  {\bibfnamefont {C.}~\bibnamefont {Mag\'en}}, \ and\ \bibinfo {author}
  {\bibfnamefont {A.}~\bibnamefont {Fert}},\ }\href {\doibase
  10.1038/ncomms3944} {\bibfield  {journal} {\bibinfo  {journal} {Nat.
  Commun.}\ }\textbf {\bibinfo {volume} {4}},\ \bibinfo {pages} {2944}
  (\bibinfo {year} {2013})}\BibitemShut {NoStop}%
\bibitem [{\citenamefont {Shen}\ \emph {et~al.}(2014)\citenamefont {Shen},
  \citenamefont {Vignale},\ and\ \citenamefont {Raimondi}}]{shen2014}%
  \BibitemOpen
  \bibfield  {author} {\bibinfo {author} {\bibfnamefont {K.}~\bibnamefont
  {Shen}}, \bibinfo {author} {\bibfnamefont {G.}~\bibnamefont {Vignale}}, \
  and\ \bibinfo {author} {\bibfnamefont {R.}~\bibnamefont {Raimondi}},\ }\href
  {\doibase 10.1103/PhysRevLett.112.096601} {\bibfield  {journal} {\bibinfo
  {journal} {Phys. Rev. Lett.}\ }\textbf {\bibinfo {volume} {112}},\ \bibinfo
  {pages} {096601} (\bibinfo {year} {2014})}\BibitemShut {NoStop}%
\bibitem [{\citenamefont {Avci}\ \emph {et~al.}(2014)\citenamefont {Avci},
  \citenamefont {Garello}, \citenamefont {Nistor}, \citenamefont {Godey},
  \citenamefont {Ballesteros}, \citenamefont {Mugarza}, \citenamefont {Barla},
  \citenamefont {Valvidares}, \citenamefont {Pellegrin}, \citenamefont {Ghosh},
  \citenamefont {Miron}, \citenamefont {Boulle}, \citenamefont {Auffret},
  \citenamefont {Gaudin},\ and\ \citenamefont {Gambardella}}]{avci2014}%
  \BibitemOpen
  \bibfield  {author} {\bibinfo {author} {\bibfnamefont {C.~O.}\ \bibnamefont
  {Avci}}, \bibinfo {author} {\bibfnamefont {K.}~\bibnamefont {Garello}},
  \bibinfo {author} {\bibfnamefont {C.}~\bibnamefont {Nistor}}, \bibinfo
  {author} {\bibfnamefont {S.}~\bibnamefont {Godey}}, \bibinfo {author}
  {\bibfnamefont {B.}~\bibnamefont {Ballesteros}}, \bibinfo {author}
  {\bibfnamefont {A.}~\bibnamefont {Mugarza}}, \bibinfo {author} {\bibfnamefont
  {A.}~\bibnamefont {Barla}}, \bibinfo {author} {\bibfnamefont
  {M.}~\bibnamefont {Valvidares}}, \bibinfo {author} {\bibfnamefont
  {E.}~\bibnamefont {Pellegrin}}, \bibinfo {author} {\bibfnamefont
  {A.}~\bibnamefont {Ghosh}}, \bibinfo {author} {\bibfnamefont {I.~M.}\
  \bibnamefont {Miron}}, \bibinfo {author} {\bibfnamefont {O.}~\bibnamefont
  {Boulle}}, \bibinfo {author} {\bibfnamefont {S.}~\bibnamefont {Auffret}},
  \bibinfo {author} {\bibfnamefont {G.}~\bibnamefont {Gaudin}}, \ and\ \bibinfo
  {author} {\bibfnamefont {P.}~\bibnamefont {Gambardella}},\ }\href {\doibase
  http://dx.doi.org/10.1103/PhysRevB.89.214419} {\bibfield  {journal} {\bibinfo
   {journal} {Phys. Rev. B}\ }\textbf {\bibinfo {volume} {89}},\ \bibinfo
  {pages} {214419} (\bibinfo {year} {2014})}\BibitemShut {NoStop}%
\bibitem [{\citenamefont {Schulz}\ \emph {et~al.}(2015)\citenamefont {Schulz},
  \citenamefont {Alejos}, \citenamefont {Martinez}, \citenamefont {Hals},
  \citenamefont {Garcia}, \citenamefont {Vila}, \citenamefont {Lee},
  \citenamefont {Conte}, \citenamefont {Karnad}, \citenamefont {Moretti},
  \citenamefont {Ocker}, \citenamefont {Ravelosona}, \citenamefont {Brataas},\
  and\ \citenamefont {Kl\"{a}ui}}]{schulz2015}%
  \BibitemOpen
  \bibfield  {author} {\bibinfo {author} {\bibfnamefont {T.}~\bibnamefont
  {Schulz}}, \bibinfo {author} {\bibfnamefont {O.}~\bibnamefont {Alejos}},
  \bibinfo {author} {\bibfnamefont {E.}~\bibnamefont {Martinez}}, \bibinfo
  {author} {\bibfnamefont {K.~M.~D.}\ \bibnamefont {Hals}}, \bibinfo {author}
  {\bibfnamefont {K.}~\bibnamefont {Garcia}}, \bibinfo {author} {\bibfnamefont
  {L.}~\bibnamefont {Vila}}, \bibinfo {author} {\bibfnamefont {K.}~\bibnamefont
  {Lee}}, \bibinfo {author} {\bibfnamefont {R.~L.}\ \bibnamefont {Conte}},
  \bibinfo {author} {\bibfnamefont {G.~V.}\ \bibnamefont {Karnad}}, \bibinfo
  {author} {\bibfnamefont {S.}~\bibnamefont {Moretti}}, \bibinfo {author}
  {\bibfnamefont {B.}~\bibnamefont {Ocker}}, \bibinfo {author} {\bibfnamefont
  {D.}~\bibnamefont {Ravelosona}}, \bibinfo {author} {\bibfnamefont
  {A.}~\bibnamefont {Brataas}}, \ and\ \bibinfo {author} {\bibfnamefont
  {M.}~\bibnamefont {Kl\"{a}ui}},\ }\href {\doibase
  http://dx.doi.org/10.1063/1.4931429} {\bibfield  {journal} {\bibinfo
  {journal} {Appl. Phys. Lett.}\ }\textbf {\bibinfo {volume} {107}},\ \bibinfo
  {pages} {122405} (\bibinfo {year} {2015})}\BibitemShut {NoStop}%
\bibitem [{\citenamefont {Gmitra}\ \emph {et~al.}(2013)\citenamefont {Gmitra},
  \citenamefont {Matos-Abiague}, \citenamefont {Draxl},\ and\ \citenamefont
  {Fabian}}]{gmitra2013}%
  \BibitemOpen
  \bibfield  {author} {\bibinfo {author} {\bibfnamefont {M.}~\bibnamefont
  {Gmitra}}, \bibinfo {author} {\bibfnamefont {A.}~\bibnamefont
  {Matos-Abiague}}, \bibinfo {author} {\bibfnamefont {C.}~\bibnamefont
  {Draxl}}, \ and\ \bibinfo {author} {\bibfnamefont {J.}~\bibnamefont
  {Fabian}},\ }\href {\doibase
  http://dx.doi.org/10.1103/PhysRevLett.111.036603} {\bibfield  {journal}
  {\bibinfo  {journal} {Phys. Rev. Lett.}\ }\textbf {\bibinfo {volume} {111}},\
  \bibinfo {pages} {036603} (\bibinfo {year} {2013})}\BibitemShut {NoStop}%
\bibitem [{\citenamefont {Hupfauer}\ \emph {et~al.}(2015)\citenamefont
  {Hupfauer}, \citenamefont {Matos-Abiague}, \citenamefont {Gmitra},
  \citenamefont {Schiller}, \citenamefont {Loher}, \citenamefont {Bougeard},
  \citenamefont {Back}, \citenamefont {Fabian},\ and\ \citenamefont
  {Weiss}}]{hupfauer2015}%
  \BibitemOpen
  \bibfield  {author} {\bibinfo {author} {\bibfnamefont {T.}~\bibnamefont
  {Hupfauer}}, \bibinfo {author} {\bibfnamefont {A.}~\bibnamefont
  {Matos-Abiague}}, \bibinfo {author} {\bibfnamefont {M.}~\bibnamefont
  {Gmitra}}, \bibinfo {author} {\bibfnamefont {F.}~\bibnamefont {Schiller}},
  \bibinfo {author} {\bibfnamefont {J.}~\bibnamefont {Loher}}, \bibinfo
  {author} {\bibfnamefont {D.}~\bibnamefont {Bougeard}}, \bibinfo {author}
  {\bibfnamefont {C.~H.}\ \bibnamefont {Back}}, \bibinfo {author}
  {\bibfnamefont {J.}~\bibnamefont {Fabian}}, \ and\ \bibinfo {author}
  {\bibfnamefont {D.}~\bibnamefont {Weiss}},\ }\href {\doibase
  DOI:10.1038/ncomms8374} {\bibfield  {journal} {\bibinfo  {journal} {Nat.
  Comm.}\ }\textbf {\bibinfo {volume} {6}},\ \bibinfo {pages} {7374} (\bibinfo
  {year} {2015})}\BibitemShut {NoStop}%
\bibitem [{\citenamefont {Shikin}\ \emph {et~al.}(2008)\citenamefont {Shikin},
  \citenamefont {Varykhalov}, \citenamefont {Prudnikova}, \citenamefont
  {Usachov}, \citenamefont {Adamchuk}, \citenamefont {Yamada}, \citenamefont
  {Riley},\ and\ \citenamefont {Rader}}]{shikin2008}%
  \BibitemOpen
  \bibfield  {author} {\bibinfo {author} {\bibfnamefont {A.~M.}\ \bibnamefont
  {Shikin}}, \bibinfo {author} {\bibfnamefont {A.}~\bibnamefont {Varykhalov}},
  \bibinfo {author} {\bibfnamefont {G.~V.}\ \bibnamefont {Prudnikova}},
  \bibinfo {author} {\bibfnamefont {D.}~\bibnamefont {Usachov}}, \bibinfo
  {author} {\bibfnamefont {V.~K.}\ \bibnamefont {Adamchuk}}, \bibinfo {author}
  {\bibfnamefont {Y.}~\bibnamefont {Yamada}}, \bibinfo {author} {\bibfnamefont
  {J.~D.}\ \bibnamefont {Riley}}, \ and\ \bibinfo {author} {\bibfnamefont
  {O.}~\bibnamefont {Rader}},\ }\href {\doibase 10.1103/PhysRevLett.100.057601}
  {\bibfield  {journal} {\bibinfo  {journal} {Phys. Rev. Lett.}\ }\textbf
  {\bibinfo {volume} {100}},\ \bibinfo {pages} {057601} (\bibinfo {year}
  {2008})}\BibitemShut {NoStop}%
\bibitem [{\citenamefont {Rybkin}\ \emph {et~al.}(2010)\citenamefont {Rybkin},
  \citenamefont {Shikin}, \citenamefont {Adamchuk}, \citenamefont {Marchenko},
  \citenamefont {Biswas}, \citenamefont {Varykhalov},\ and\ \citenamefont
  {Rader}}]{rybkin2010}%
  \BibitemOpen
  \bibfield  {author} {\bibinfo {author} {\bibfnamefont {A.~G.}\ \bibnamefont
  {Rybkin}}, \bibinfo {author} {\bibfnamefont {A.~M.}\ \bibnamefont {Shikin}},
  \bibinfo {author} {\bibfnamefont {V.~K.}\ \bibnamefont {Adamchuk}}, \bibinfo
  {author} {\bibfnamefont {D.}~\bibnamefont {Marchenko}}, \bibinfo {author}
  {\bibfnamefont {C.}~\bibnamefont {Biswas}}, \bibinfo {author} {\bibfnamefont
  {A.}~\bibnamefont {Varykhalov}}, \ and\ \bibinfo {author} {\bibfnamefont
  {O.}~\bibnamefont {Rader}},\ }\href {\doibase 10.1103/PhysRevB.82.233403}
  {\bibfield  {journal} {\bibinfo  {journal} {Phys. Rev. B}\ }\textbf {\bibinfo
  {volume} {82}},\ \bibinfo {pages} {233403} (\bibinfo {year}
  {2010})}\BibitemShut {NoStop}%
\bibitem [{\citenamefont {Valenzuela}\ and\ \citenamefont
  {Tinkham}(2006)}]{valenzuela2006}%
  \BibitemOpen
  \bibfield  {author} {\bibinfo {author} {\bibfnamefont {S.~O.}\ \bibnamefont
  {Valenzuela}}\ and\ \bibinfo {author} {\bibfnamefont {M.}~\bibnamefont
  {Tinkham}},\ }\href {\doibase 10.1038/nature04937} {\bibfield  {journal}
  {\bibinfo  {journal} {Nature}\ }\textbf {\bibinfo {volume} {442}},\ \bibinfo
  {pages} {176} (\bibinfo {year} {2006})}\BibitemShut {NoStop}%
\bibitem [{\citenamefont {Shiomi}\ \emph {et~al.}(2014)\citenamefont {Shiomi},
  \citenamefont {Nomura}, \citenamefont {Kajiwara}, \citenamefont {Eto},
  \citenamefont {Novak}, \citenamefont {Segawa}, \citenamefont {Ando},\ and\
  \citenamefont {Saitoh}}]{shiomi2014}%
  \BibitemOpen
  \bibfield  {author} {\bibinfo {author} {\bibfnamefont {Y.}~\bibnamefont
  {Shiomi}}, \bibinfo {author} {\bibfnamefont {K.}~\bibnamefont {Nomura}},
  \bibinfo {author} {\bibfnamefont {Y.}~\bibnamefont {Kajiwara}}, \bibinfo
  {author} {\bibfnamefont {K.}~\bibnamefont {Eto}}, \bibinfo {author}
  {\bibfnamefont {M.}~\bibnamefont {Novak}}, \bibinfo {author} {\bibfnamefont
  {K.}~\bibnamefont {Segawa}}, \bibinfo {author} {\bibfnamefont
  {Y.}~\bibnamefont {Ando}}, \ and\ \bibinfo {author} {\bibfnamefont
  {E.}~\bibnamefont {Saitoh}},\ }\href {\doibase
  http://dx.doi.org/10.1103/PhysRevLett.113.196601} {\bibfield  {journal}
  {\bibinfo  {journal} {Phys. Rev. Lett.}\ }\textbf {\bibinfo {volume} {113}},\
  \bibinfo {pages} {196601} (\bibinfo {year} {2014})}\BibitemShut {NoStop}%
\bibitem [{\citenamefont {Mellnik}\ \emph {et~al.}(2014)\citenamefont
  {Mellnik}, \citenamefont {Lee}, \citenamefont {Richardella}, \citenamefont
  {Grab}, \citenamefont {Mintun}, \citenamefont {Fischer}, \citenamefont
  {Vaezi}, \citenamefont {Manchon}, \citenamefont {Kim}, \citenamefont
  {Samarth},\ and\ \citenamefont {Ralph}}]{mellnik2014}%
  \BibitemOpen
  \bibfield  {author} {\bibinfo {author} {\bibfnamefont {A.~R.}\ \bibnamefont
  {Mellnik}}, \bibinfo {author} {\bibfnamefont {J.~S.}\ \bibnamefont {Lee}},
  \bibinfo {author} {\bibfnamefont {A.}~\bibnamefont {Richardella}}, \bibinfo
  {author} {\bibfnamefont {J.~L.}\ \bibnamefont {Grab}}, \bibinfo {author}
  {\bibfnamefont {P.~J.}\ \bibnamefont {Mintun}}, \bibinfo {author}
  {\bibfnamefont {M.~H.}\ \bibnamefont {Fischer}}, \bibinfo {author}
  {\bibfnamefont {A.}~\bibnamefont {Vaezi}}, \bibinfo {author} {\bibfnamefont
  {A.}~\bibnamefont {Manchon}}, \bibinfo {author} {\bibfnamefont {E.-A.}\
  \bibnamefont {Kim}}, \bibinfo {author} {\bibfnamefont {N.}~\bibnamefont
  {Samarth}}, \ and\ \bibinfo {author} {\bibfnamefont {D.~C.}\ \bibnamefont
  {Ralph}},\ }\href {\doibase doi:10.1038/nature13534} {\bibfield  {journal}
  {\bibinfo  {journal} {Nature}\ }\textbf {\bibinfo {volume} {511}},\ \bibinfo
  {pages} {449} (\bibinfo {year} {2014})}\BibitemShut {NoStop}%
\bibitem [{\citenamefont {Raimondi}\ and\ \citenamefont
  {Schwab}(2009)}]{raimondi2009}%
  \BibitemOpen
  \bibfield  {author} {\bibinfo {author} {\bibfnamefont {R.}~\bibnamefont
  {Raimondi}}\ and\ \bibinfo {author} {\bibfnamefont {P.}~\bibnamefont
  {Schwab}},\ }\href {\doibase http://dx.doi.org/10.1209/0295-5075/87/37008}
  {\bibfield  {journal} {\bibinfo  {journal} {Europhys. Lett.}\ }\textbf
  {\bibinfo {volume} {87}},\ \bibinfo {pages} {37008} (\bibinfo {year}
  {2009})}\BibitemShut {NoStop}%
\bibitem [{\citenamefont {Tserkovnyak}\ \emph {et~al.}(2008)\citenamefont
  {Tserkovnyak}, \citenamefont {Brataas},\ and\ \citenamefont
  {Bauer}}]{tserkovnyak2008_1}%
  \BibitemOpen
  \bibfield  {author} {\bibinfo {author} {\bibfnamefont {Y.}~\bibnamefont
  {Tserkovnyak}}, \bibinfo {author} {\bibfnamefont {A.}~\bibnamefont
  {Brataas}}, \ and\ \bibinfo {author} {\bibfnamefont {G.~E.}\ \bibnamefont
  {Bauer}},\ }\href {\doibase http://dx.doi.org/10.1016/j.jmmm.2007.12.012}
  {\bibfield  {journal} {\bibinfo  {journal} {J. Magn. Magn. Mater.}\ }\textbf
  {\bibinfo {volume} {320}},\ \bibinfo {pages} {1282 } (\bibinfo {year}
  {2008})}\BibitemShut {NoStop}%
\bibitem [{\citenamefont {Tserkovnyak}\ and\ \citenamefont
  {Mecklenburg}(2008)}]{tserkovnyak2008_2}%
  \BibitemOpen
  \bibfield  {author} {\bibinfo {author} {\bibfnamefont {Y.}~\bibnamefont
  {Tserkovnyak}}\ and\ \bibinfo {author} {\bibfnamefont {M.}~\bibnamefont
  {Mecklenburg}},\ }\href {\doibase 10.1103/PhysRevB.77.134407} {\bibfield
  {journal} {\bibinfo  {journal} {Phys. Rev. B}\ }\textbf {\bibinfo {volume}
  {77}},\ \bibinfo {pages} {134407} (\bibinfo {year} {2008})}\BibitemShut
  {NoStop}%
\bibitem [{\citenamefont {Vlaminck}\ and\ \citenamefont
  {Bailleul}(2008)}]{vlaminck2008}%
  \BibitemOpen
  \bibfield  {author} {\bibinfo {author} {\bibfnamefont {V.}~\bibnamefont
  {Vlaminck}}\ and\ \bibinfo {author} {\bibfnamefont {M.}~\bibnamefont
  {Bailleul}},\ }\href {\doibase DOI: 10.1126/science.1162843} {\bibfield
  {journal} {\bibinfo  {journal} {Science}\ }\textbf {\bibinfo {volume}
  {322}},\ \bibinfo {pages} {410} (\bibinfo {year} {2008})}\BibitemShut
  {NoStop}%
\bibitem [{\citenamefont {Kajiwara}\ \emph {et~al.}(2010)\citenamefont
  {Kajiwara}, \citenamefont {Harii}, \citenamefont {Takahashi}, \citenamefont
  {Ohe}, \citenamefont {Uchida}, \citenamefont {Mizuguchi}, \citenamefont
  {Umezawa}, \citenamefont {Kawai}, \citenamefont {Ando}, \citenamefont
  {Takanashi}, \citenamefont {Maekawa},\ and\ \citenamefont
  {Saitoh}}]{kajiwara2010}%
  \BibitemOpen
  \bibfield  {author} {\bibinfo {author} {\bibfnamefont {Y.}~\bibnamefont
  {Kajiwara}}, \bibinfo {author} {\bibfnamefont {K.}~\bibnamefont {Harii}},
  \bibinfo {author} {\bibfnamefont {S.}~\bibnamefont {Takahashi}}, \bibinfo
  {author} {\bibfnamefont {J.}~\bibnamefont {Ohe}}, \bibinfo {author}
  {\bibfnamefont {K.}~\bibnamefont {Uchida}}, \bibinfo {author} {\bibfnamefont
  {M.}~\bibnamefont {Mizuguchi}}, \bibinfo {author} {\bibfnamefont
  {H.}~\bibnamefont {Umezawa}}, \bibinfo {author} {\bibfnamefont
  {H.}~\bibnamefont {Kawai}}, \bibinfo {author} {\bibfnamefont
  {K.}~\bibnamefont {Ando}}, \bibinfo {author} {\bibfnamefont {K.}~\bibnamefont
  {Takanashi}}, \bibinfo {author} {\bibfnamefont {S.}~\bibnamefont {Maekawa}},
  \ and\ \bibinfo {author} {\bibfnamefont {E.}~\bibnamefont {Saitoh}},\ }\href
  {\doibase doi:10.1038/nature08876} {\bibfield  {journal} {\bibinfo  {journal}
  {Nature}\ }\textbf {\bibinfo {volume} {464}},\ \bibinfo {pages} {262}
  (\bibinfo {year} {2010})}\BibitemShut {NoStop}%
\bibitem [{\citenamefont {Gorini}\ \emph {et~al.}(2010)\citenamefont {Gorini},
  \citenamefont {Schwab}, \citenamefont {Raimondi},\ and\ \citenamefont
  {Shelankov}}]{gorini2010}%
  \BibitemOpen
  \bibfield  {author} {\bibinfo {author} {\bibfnamefont {C.}~\bibnamefont
  {Gorini}}, \bibinfo {author} {\bibfnamefont {P.}~\bibnamefont {Schwab}},
  \bibinfo {author} {\bibfnamefont {R.}~\bibnamefont {Raimondi}}, \ and\
  \bibinfo {author} {\bibfnamefont {A.~L.}\ \bibnamefont {Shelankov}},\ }\href
  {\doibase 10.1103/PhysRevB.82.195316} {\bibfield  {journal} {\bibinfo
  {journal} {Phys. Rev. B}\ }\textbf {\bibinfo {volume} {82}},\ \bibinfo
  {pages} {195316} (\bibinfo {year} {2010})}\BibitemShut {NoStop}%
\bibitem [{\citenamefont {T\"olle}\ \emph {et~al.}(2014)\citenamefont
  {T\"olle}, \citenamefont {Gorini},\ and\ \citenamefont
  {Eckern}}]{toelle2014}%
  \BibitemOpen
  \bibfield  {author} {\bibinfo {author} {\bibfnamefont {S.}~\bibnamefont
  {T\"olle}}, \bibinfo {author} {\bibfnamefont {C.}~\bibnamefont {Gorini}}, \
  and\ \bibinfo {author} {\bibfnamefont {U.}~\bibnamefont {Eckern}},\ }\href
  {\doibase 10.1103/PhysRevB.90.235117} {\bibfield  {journal} {\bibinfo
  {journal} {Phys. Rev. B}\ }\textbf {\bibinfo {volume} {90}},\ \bibinfo
  {pages} {235117} (\bibinfo {year} {2014})}\BibitemShut {NoStop}%
\bibitem [{\citenamefont {Wang}\ \emph {et~al.}(2013)\citenamefont {Wang},
  \citenamefont {Xiao}, \citenamefont {Manchon},\ and\ \citenamefont
  {Maekawa}}]{wang2013}%
  \BibitemOpen
  \bibfield  {author} {\bibinfo {author} {\bibfnamefont {X.}~\bibnamefont
  {Wang}}, \bibinfo {author} {\bibfnamefont {J.}~\bibnamefont {Xiao}}, \bibinfo
  {author} {\bibfnamefont {A.}~\bibnamefont {Manchon}}, \ and\ \bibinfo
  {author} {\bibfnamefont {S.}~\bibnamefont {Maekawa}},\ }\href {\doibase
  10.1103/PhysRevB.87.081407} {\bibfield  {journal} {\bibinfo  {journal} {Phys.
  Rev. B}\ }\textbf {\bibinfo {volume} {87}},\ \bibinfo {pages} {081407}
  (\bibinfo {year} {2013})}\BibitemShut {NoStop}%
\bibitem [{\citenamefont {Haney}\ \emph {et~al.}(2013)\citenamefont {Haney},
  \citenamefont {Lee}, \citenamefont {Lee}, \citenamefont {Manchon},\ and\
  \citenamefont {Stiles}}]{haney2013}%
  \BibitemOpen
  \bibfield  {author} {\bibinfo {author} {\bibfnamefont {P.~M.}\ \bibnamefont
  {Haney}}, \bibinfo {author} {\bibfnamefont {H.-W.}\ \bibnamefont {Lee}},
  \bibinfo {author} {\bibfnamefont {K.-J.}\ \bibnamefont {Lee}}, \bibinfo
  {author} {\bibfnamefont {A.}~\bibnamefont {Manchon}}, \ and\ \bibinfo
  {author} {\bibfnamefont {M.~D.}\ \bibnamefont {Stiles}},\ }\href {\doibase
  10.1103/PhysRevB.87.174411} {\bibfield  {journal} {\bibinfo  {journal} {Phys.
  Rev. B}\ }\textbf {\bibinfo {volume} {87}},\ \bibinfo {pages} {174411}
  (\bibinfo {year} {2013})}\BibitemShut {NoStop}%
\bibitem [{\citenamefont {Chen}\ and\ \citenamefont {Zhang}(2015)}]{chen2015}%
  \BibitemOpen
  \bibfield  {author} {\bibinfo {author} {\bibfnamefont {K.}~\bibnamefont
  {Chen}}\ and\ \bibinfo {author} {\bibfnamefont {S.}~\bibnamefont {Zhang}},\
  }\href {\doibase 10.1103/PhysRevLett.114.126602} {\bibfield  {journal}
  {\bibinfo  {journal} {Phys. Rev. Lett.}\ }\textbf {\bibinfo {volume} {114}},\
  \bibinfo {pages} {126602} (\bibinfo {year} {2015})}\BibitemShut {NoStop}%
\bibitem [{\citenamefont {Amin}\ and\ \citenamefont
  {Stiles}(2016{\natexlab{a}})}]{amin2016a}%
  \BibitemOpen
  \bibfield  {author} {\bibinfo {author} {\bibfnamefont {V.~P.}\ \bibnamefont
  {Amin}}\ and\ \bibinfo {author} {\bibfnamefont {M.~D.}\ \bibnamefont
  {Stiles}},\ }\href {\doibase 10.1103/PhysRevB.94.104419} {\bibfield
  {journal} {\bibinfo  {journal} {Phys. Rev. B}\ }\textbf {\bibinfo {volume}
  {94}},\ \bibinfo {pages} {104419} (\bibinfo {year}
  {2016}{\natexlab{a}})}\BibitemShut {NoStop}%
\bibitem [{\citenamefont {Amin}\ and\ \citenamefont
  {Stiles}(2016{\natexlab{b}})}]{amin2016b}%
  \BibitemOpen
  \bibfield  {author} {\bibinfo {author} {\bibfnamefont {V.~P.}\ \bibnamefont
  {Amin}}\ and\ \bibinfo {author} {\bibfnamefont {M.~D.}\ \bibnamefont
  {Stiles}},\ }\href {\doibase 10.1103/PhysRevB.94.104420} {\bibfield
  {journal} {\bibinfo  {journal} {Phys. Rev. B}\ }\textbf {\bibinfo {volume}
  {94}},\ \bibinfo {pages} {104420} (\bibinfo {year}
  {2016}{\natexlab{b}})}\BibitemShut {NoStop}%
\bibitem [{\citenamefont {Antel}\ \emph {et~al.}(1999)\citenamefont {Antel},
  \citenamefont {Schwickert}, \citenamefont {Lin}, \citenamefont {O'Brien},\
  and\ \citenamefont {Harp}}]{antel1999}%
  \BibitemOpen
  \bibfield  {author} {\bibinfo {author} {\bibfnamefont {W.~J.}\ \bibnamefont
  {Antel}}, \bibinfo {author} {\bibfnamefont {M.~M.}\ \bibnamefont
  {Schwickert}}, \bibinfo {author} {\bibfnamefont {T.}~\bibnamefont {Lin}},
  \bibinfo {author} {\bibfnamefont {W.~L.}\ \bibnamefont {O'Brien}}, \ and\
  \bibinfo {author} {\bibfnamefont {G.~R.}\ \bibnamefont {Harp}},\ }\href
  {\doibase 10.1103/PhysRevB.60.12933} {\bibfield  {journal} {\bibinfo
  {journal} {Phys. Rev. B}\ }\textbf {\bibinfo {volume} {60}},\ \bibinfo
  {pages} {12933} (\bibinfo {year} {1999})}\BibitemShut {NoStop}%
\bibitem [{\citenamefont {Wilhelm}\ \emph {et~al.}(2000)\citenamefont
  {Wilhelm}, \citenamefont {Poulopoulos}, \citenamefont {Ceballos},
  \citenamefont {Wende}, \citenamefont {Baberschke}, \citenamefont
  {Srivastava}, \citenamefont {Benea}, \citenamefont {Ebert}, \citenamefont
  {Angelakeris}, \citenamefont {Flevaris}, \citenamefont {Niarchos},
  \citenamefont {Rogalev},\ and\ \citenamefont {Brookes}}]{wilhelm2000}%
  \BibitemOpen
  \bibfield  {author} {\bibinfo {author} {\bibfnamefont {F.}~\bibnamefont
  {Wilhelm}}, \bibinfo {author} {\bibfnamefont {P.}~\bibnamefont
  {Poulopoulos}}, \bibinfo {author} {\bibfnamefont {G.}~\bibnamefont
  {Ceballos}}, \bibinfo {author} {\bibfnamefont {H.}~\bibnamefont {Wende}},
  \bibinfo {author} {\bibfnamefont {K.}~\bibnamefont {Baberschke}}, \bibinfo
  {author} {\bibfnamefont {P.}~\bibnamefont {Srivastava}}, \bibinfo {author}
  {\bibfnamefont {D.}~\bibnamefont {Benea}}, \bibinfo {author} {\bibfnamefont
  {H.}~\bibnamefont {Ebert}}, \bibinfo {author} {\bibfnamefont
  {M.}~\bibnamefont {Angelakeris}}, \bibinfo {author} {\bibfnamefont {N.~K.}\
  \bibnamefont {Flevaris}}, \bibinfo {author} {\bibfnamefont {D.}~\bibnamefont
  {Niarchos}}, \bibinfo {author} {\bibfnamefont {A.}~\bibnamefont {Rogalev}}, \
  and\ \bibinfo {author} {\bibfnamefont {N.~B.}\ \bibnamefont {Brookes}},\
  }\href {\doibase 10.1103/PhysRevLett.85.413} {\bibfield  {journal} {\bibinfo
  {journal} {Phys. Rev. Lett.}\ }\textbf {\bibinfo {volume} {85}},\ \bibinfo
  {pages} {413} (\bibinfo {year} {2000})}\BibitemShut {NoStop}%
\bibitem [{\citenamefont {Guo}\ \emph {et~al.}(2014)\citenamefont {Guo},
  \citenamefont {Niu},\ and\ \citenamefont {Nagaosa}}]{guo2014}%
  \BibitemOpen
  \bibfield  {author} {\bibinfo {author} {\bibfnamefont {G.~Y.}\ \bibnamefont
  {Guo}}, \bibinfo {author} {\bibfnamefont {Q.}~\bibnamefont {Niu}}, \ and\
  \bibinfo {author} {\bibfnamefont {N.}~\bibnamefont {Nagaosa}},\ }\href
  {\doibase 10.1103/PhysRevB.89.214406} {\bibfield  {journal} {\bibinfo
  {journal} {Phys. Rev. B}\ }\textbf {\bibinfo {volume} {89}},\ \bibinfo
  {pages} {214406} (\bibinfo {year} {2014})}\BibitemShut {NoStop}%
\bibitem [{\citenamefont {Ast}\ \emph {et~al.}(2007)\citenamefont {Ast},
  \citenamefont {Henk}, \citenamefont {Ernst}, \citenamefont {Moreschini},
  \citenamefont {Falub}, \citenamefont {Pacil\'e}, \citenamefont {Bruno},
  \citenamefont {Kern},\ and\ \citenamefont {Grioni}}]{ast2007}%
  \BibitemOpen
  \bibfield  {author} {\bibinfo {author} {\bibfnamefont {C.~R.}\ \bibnamefont
  {Ast}}, \bibinfo {author} {\bibfnamefont {J.}~\bibnamefont {Henk}}, \bibinfo
  {author} {\bibfnamefont {A.}~\bibnamefont {Ernst}}, \bibinfo {author}
  {\bibfnamefont {L.}~\bibnamefont {Moreschini}}, \bibinfo {author}
  {\bibfnamefont {M.~C.}\ \bibnamefont {Falub}}, \bibinfo {author}
  {\bibfnamefont {D.}~\bibnamefont {Pacil\'e}}, \bibinfo {author}
  {\bibfnamefont {P.}~\bibnamefont {Bruno}}, \bibinfo {author} {\bibfnamefont
  {K.}~\bibnamefont {Kern}}, \ and\ \bibinfo {author} {\bibfnamefont
  {M.}~\bibnamefont {Grioni}},\ }\href {\doibase 10.1103/PhysRevLett.98.186807}
  {\bibfield  {journal} {\bibinfo  {journal} {Phys. Rev. Lett.}\ }\textbf
  {\bibinfo {volume} {98}},\ \bibinfo {pages} {186807} (\bibinfo {year}
  {2007})}\BibitemShut {NoStop}%
\bibitem [{\citenamefont {Gorini}\ \emph {et~al.}(2017)\citenamefont {Gorini},
  \citenamefont {Maleki}, \citenamefont {Shen}, \citenamefont {Tokatly},
  \citenamefont {Vignale},\ and\ \citenamefont {Raimondi}}]{gorini2016}%
  \BibitemOpen
  \bibfield  {author} {\bibinfo {author} {\bibfnamefont {C.}~\bibnamefont
  {Gorini}}, \bibinfo {author} {\bibfnamefont {A.}~\bibnamefont {Maleki}},
  \bibinfo {author} {\bibfnamefont {K.}~\bibnamefont {Shen}}, \bibinfo {author}
  {\bibfnamefont {I.~V.}\ \bibnamefont {Tokatly}}, \bibinfo {author}
  {\bibfnamefont {G.}~\bibnamefont {Vignale}}, \ and\ \bibinfo {author}
  {\bibfnamefont {R.}~\bibnamefont {Raimondi}},\ }\href
  {https://arxiv.org/abs/1702.04887} {\bibfield  {journal} {\bibinfo  {journal}
  {arXiv:1702.04887}\ } (\bibinfo {year} {2017})}\BibitemShut {NoStop}%
\bibitem [{Note2()}]{Note2}%
  \BibitemOpen
  \bibinfo {note} {Lower indices refer to the spatial component, whereas upper
  indices represent the polarization. When the spin current is written in
  boldface this means a vector consisting of the three components which are not
  marked as index.}\BibitemShut {Stop}%
\bibitem [{\citenamefont {Knoester}\ \emph {et~al.}(2014)\citenamefont
  {Knoester}, \citenamefont {Sinova},\ and\ \citenamefont
  {Duine}}]{knoester2014}%
  \BibitemOpen
  \bibfield  {author} {\bibinfo {author} {\bibfnamefont {M.~E.}\ \bibnamefont
  {Knoester}}, \bibinfo {author} {\bibfnamefont {J.}~\bibnamefont {Sinova}}, \
  and\ \bibinfo {author} {\bibfnamefont {R.~A.}\ \bibnamefont {Duine}},\ }\href
  {\doibase 10.1103/PhysRevB.89.064425} {\bibfield  {journal} {\bibinfo
  {journal} {Phys. Rev. B}\ }\textbf {\bibinfo {volume} {89}},\ \bibinfo
  {pages} {064425} (\bibinfo {year} {2014})}\BibitemShut {NoStop}%
\bibitem [{\citenamefont {Stern}(1992)}]{stern1992}%
  \BibitemOpen
  \bibfield  {author} {\bibinfo {author} {\bibfnamefont {A.}~\bibnamefont
  {Stern}},\ }\href {\doibase 10.1103/PhysRevLett.68.1022} {\bibfield
  {journal} {\bibinfo  {journal} {Phys. Rev. Lett.}\ }\textbf {\bibinfo
  {volume} {68}},\ \bibinfo {pages} {1022} (\bibinfo {year}
  {1992})}\BibitemShut {NoStop}%
\bibitem [{\citenamefont {Barnes}\ and\ \citenamefont
  {Maekawa}(2007)}]{barnes2007}%
  \BibitemOpen
  \bibfield  {author} {\bibinfo {author} {\bibfnamefont {S.~E.}\ \bibnamefont
  {Barnes}}\ and\ \bibinfo {author} {\bibfnamefont {S.}~\bibnamefont
  {Maekawa}},\ }\href {\doibase 10.1103/PhysRevLett.98.246601} {\bibfield
  {journal} {\bibinfo  {journal} {Phys. Rev. Lett.}\ }\textbf {\bibinfo
  {volume} {98}},\ \bibinfo {pages} {246601} (\bibinfo {year}
  {2007})}\BibitemShut {NoStop}%
\bibitem [{\citenamefont {Duine}(2008)}]{duine2008}%
  \BibitemOpen
  \bibfield  {author} {\bibinfo {author} {\bibfnamefont {R.~A.}\ \bibnamefont
  {Duine}},\ }\href {\doibase 10.1103/PhysRevB.77.014409} {\bibfield  {journal}
  {\bibinfo  {journal} {Phys. Rev. B}\ }\textbf {\bibinfo {volume} {77}},\
  \bibinfo {pages} {014409} (\bibinfo {year} {2008})}\BibitemShut {NoStop}%
\bibitem [{\citenamefont {Tatara}\ \emph {et~al.}(2013)\citenamefont {Tatara},
  \citenamefont {Nakabayashi},\ and\ \citenamefont {Lee}}]{tatara2013}%
  \BibitemOpen
  \bibfield  {author} {\bibinfo {author} {\bibfnamefont {G.}~\bibnamefont
  {Tatara}}, \bibinfo {author} {\bibfnamefont {N.}~\bibnamefont {Nakabayashi}},
  \ and\ \bibinfo {author} {\bibfnamefont {K.-J.}\ \bibnamefont {Lee}},\ }\href
  {\doibase 10.1103/PhysRevB.87.054403} {\bibfield  {journal} {\bibinfo
  {journal} {Phys. Rev. B}\ }\textbf {\bibinfo {volume} {87}},\ \bibinfo
  {pages} {054403} (\bibinfo {year} {2013})}\BibitemShut {NoStop}%
\bibitem [{\citenamefont {Yamane}\ \emph {et~al.}(2013)\citenamefont {Yamane},
  \citenamefont {Ieda},\ and\ \citenamefont {Maekawa}}]{yamane2013}%
  \BibitemOpen
  \bibfield  {author} {\bibinfo {author} {\bibfnamefont {Y.}~\bibnamefont
  {Yamane}}, \bibinfo {author} {\bibfnamefont {J.}~\bibnamefont {Ieda}}, \ and\
  \bibinfo {author} {\bibfnamefont {S.}~\bibnamefont {Maekawa}},\ }\href
  {\doibase 10.1103/PhysRevB.88.014430} {\bibfield  {journal} {\bibinfo
  {journal} {Phys. Rev. B}\ }\textbf {\bibinfo {volume} {88}},\ \bibinfo
  {pages} {014430} (\bibinfo {year} {2013})}\BibitemShut {NoStop}%
\bibitem [{\citenamefont {Raimondi}\ \emph {et~al.}(2012)\citenamefont
  {Raimondi}, \citenamefont {Schwab}, \citenamefont {Gorini},\ and\
  \citenamefont {Vignale}}]{raimondi2012}%
  \BibitemOpen
  \bibfield  {author} {\bibinfo {author} {\bibfnamefont {R.}~\bibnamefont
  {Raimondi}}, \bibinfo {author} {\bibfnamefont {P.}~\bibnamefont {Schwab}},
  \bibinfo {author} {\bibfnamefont {C.}~\bibnamefont {Gorini}}, \ and\ \bibinfo
  {author} {\bibfnamefont {G.}~\bibnamefont {Vignale}},\ }\href {\doibase
  10.1002/andp.201100253} {\bibfield  {journal} {\bibinfo  {journal} {Ann.
  Phys. (Berlin)}\ }\textbf {\bibinfo {volume} {524}},\ \bibinfo {pages} {153}
  (\bibinfo {year} {2012})}\BibitemShut {NoStop}%
\bibitem [{\citenamefont {Schwab}\ \emph {et~al.}(2010)\citenamefont {Schwab},
  \citenamefont {Raimondi},\ and\ \citenamefont {Gorini}}]{schwab2010}%
  \BibitemOpen
  \bibfield  {author} {\bibinfo {author} {\bibfnamefont {P.}~\bibnamefont
  {Schwab}}, \bibinfo {author} {\bibfnamefont {R.}~\bibnamefont {Raimondi}}, \
  and\ \bibinfo {author} {\bibfnamefont {C.}~\bibnamefont {Gorini}},\ }\href
  {\doibase 10.1209/0295-5075/90/67004} {\bibfield  {journal} {\bibinfo
  {journal} {Europhys. Lett.}\ }\textbf {\bibinfo {volume} {90}},\ \bibinfo
  {pages} {67004} (\bibinfo {year} {2010})}\BibitemShut {NoStop}%
\bibitem [{Note3()}]{Note3}%
  \BibitemOpen
  \bibinfo {note} {Experimentally, the setup employed to excite the
  magnetization dynamics has to be carefully chosen, see the discussion in
  Ref.~[\protect \rev@citealpnum {obstbaum2014}].}\BibitemShut {Stop}%
\bibitem [{Note4()}]{Note4}%
  \BibitemOpen
  \bibinfo {note} {{Note that in this work we do not consider the side-jump and
  skew-scattering contributions. Hence, from the first part of Eq.~(36) in
  Ref.~\protect \rev@citealpnum {raimondi2012} we obtain for $\tau _s\gg \tau
  _\protect \mathrm {DP}$, cf.\ Eq.~\protect \textup {\hbox {\mathsurround \z@
  \protect \normalfont (\ignorespaces \ref {hierarchy}\unskip \@@italiccorr
  )}}, $\sigma ^\protect \mathrm {sH} = (\tau _\protect \mathrm {DP}/\tau
  _s)\sigma ^\protect \mathrm {sH}_\protect \mathrm {int}$ which, together with
  $\sigma ^\protect \mathrm {sH}_\protect \mathrm {int} = e\tau /4\pi \tau
  _\protect \mathrm {DP}$, leads to the result given.}}\BibitemShut {Stop}%
\bibitem [{Note5()}]{Note5}%
  \BibitemOpen
  \bibinfo {note} {More precisely, parallel to ${\protect \bf n}_{\zeta
  }$.}\BibitemShut {Stop}%
\bibitem [{\citenamefont {Tserkovnyak}\ \emph {et~al.}(2005)\citenamefont
  {Tserkovnyak}, \citenamefont {Brataas}, \citenamefont {Bauer},\ and\
  \citenamefont {Halperin}}]{tserkovnyak2005}%
  \BibitemOpen
  \bibfield  {author} {\bibinfo {author} {\bibfnamefont {Y.}~\bibnamefont
  {Tserkovnyak}}, \bibinfo {author} {\bibfnamefont {A.}~\bibnamefont
  {Brataas}}, \bibinfo {author} {\bibfnamefont {G.~E.~W.}\ \bibnamefont
  {Bauer}}, \ and\ \bibinfo {author} {\bibfnamefont {B.~I.}\ \bibnamefont
  {Halperin}},\ }\href {\doibase 10.1103/RevModPhys.77.1375} {\bibfield
  {journal} {\bibinfo  {journal} {Rev. Mod. Phys.}\ }\textbf {\bibinfo {volume}
  {77}},\ \bibinfo {pages} {1375} (\bibinfo {year} {2005})}\BibitemShut
  {NoStop}%
\bibitem [{\citenamefont {Shpiro}\ \emph {et~al.}(2003)\citenamefont {Shpiro},
  \citenamefont {Levy},\ and\ \citenamefont {Zhang}}]{asaya2012}%
  \BibitemOpen
  \bibfield  {author} {\bibinfo {author} {\bibfnamefont {A.}~\bibnamefont
  {Shpiro}}, \bibinfo {author} {\bibfnamefont {P.~M.}\ \bibnamefont {Levy}}, \
  and\ \bibinfo {author} {\bibfnamefont {S.}~\bibnamefont {Zhang}},\ }\href
  {\doibase 10.1103/PhysRevB.67.104430} {\bibfield  {journal} {\bibinfo
  {journal} {Phys. Rev. B}\ }\textbf {\bibinfo {volume} {67}},\ \bibinfo
  {pages} {104430} (\bibinfo {year} {2003})}\BibitemShut {NoStop}%
\end{thebibliography}%
\end{document}